\documentclass[twocolumn,showpacs,preprintnumbers,amsmath,amssymb]{revtex4}

\usepackage{graphicx}
\usepackage{dcolumn}
\usepackage{bm}
\usepackage{epsfig}
\usepackage{amssymb}
\usepackage{natbib}

\begin{document}

\title{Effects of payoff functions and preference distributions
in an adaptive population}

\author{H. M. Yang, Y. S. Ting, and K. Y. Michael Wong}
\email{hmyang@ust.hk, phkywong@ust.hk}
\affiliation{ Department of Physics,
Hong Kong University of Science and Technology, Hong Kong,
China}

\begin{abstract}
Adaptive populations such as those
in financial markets and distributed control
can be modeled by the Minority Game.
We consider how their dynamics
depends on the agents' initial preferences of strategies,
when the agents use linear or quadratic payoff functions
to evaluate their strategies.
We find that the fluctuations of the population
making certain decisions (the volatility)
depends on the diversity of the distribution
of the initial preferences of strategies.
When the diversity decreases,
more agents tend to adapt their strategies together.
In systems with linear payoffs,
this results in dynamical transitions
from vanishing volatility to a non-vanishing one.
For low signal dimensions,
the dynamical transitions for the different signals
do not take place at the same critical diversity.
Rather, a cascade of dynamical transitions takes place
when the diversity is reduced.
In contrast, no phase transitions are found
in systems with the quadratic payoffs.
Instead, a basin boundary of attraction
separates two groups of samples
in the space of the agents' decisions.
Initial states inside this boundary
converge to small volatility,
while those outside diverge to a large one.
Furthermore, when the preference distribution
becomes more polarized,
the dynamics becomes more erratic.
All the above results are supported
by good agreement between simulations and theory.
\end{abstract}

\maketitle

\section{INTRODUCTION}

Many natural and artificial systems consist of a population of agents with coupled dynamics. Through their
mutual adaptation, they are able to exhibit interesting collective behavior. Although the individuals are
competing to maximize their own payoffs, the system is able to self-organize itself to globally efficient
states. Examples can be found in economic markets and communication networks
\cite{Anderson1988,Challet1997,Wei1995,Schweitzer2002}.

As a prototype of an adaptive population, the Minority Game (MG) considers the dynamics of buyers and sellers in
a model of the financial market in which the minority group is the winning one \cite{Challet1997}. The agents
adapt to each other through adjusting the payoffs of their strategies, wherein the payoffs summarize the
collective behavior of the population when the environment changes. Theoretical studies using the replica method
\cite{Challet2000, Marsili2000} and the generating functional \cite{Heimel2001, Coolen2001, Coolen2005}
successfully describe the statistical properties of these systems. However, since adaptation is a dynamical
process, much of the attractor behavior remains to be explored.

An important factor affecting the behavior of an adaptive population
is the dependence of the payoffs on the environment experienced by
the individual agents. The payoffs help the agents to assess the
preferences of their decisions, hence inducing them to take certain
actions when they experience a similar dynamical environment in the
future. Thus, the payoff function is crucial to the mechanism of
adaptation.
For example, the ways the agents evaluate their strategies in financial markets have a large influence on the
market behavior. There are agents who focus their attention mainly on opportunities of large profits, while
others pay equal attention to all profitable opportunities large and small. Markets with individualistic agents
may have diversified opinions about what the best strategies are.

As an illustration of how payoffs influence market behaviors,
a payoff scheme based on expected price trends
results in markets with a mixture of trend-followers
and contrarians~\cite{Marsili2001}.
The \$-Game considers payoffs
rewarding correct expectations one step forward,
giving rise to a self-sustained speculative phase~\cite{Andersen2003}.
Bubbles, crashes and intermittent behaviors are also found
in a similar extension of the MG~\cite{Giardina2003}.
A recent extension of the MG considers agents
rewarding trend-following strategies
when the winning margin is small,
and rewarding contrarian ones
otherwise~\cite{de Martino2004,Tedeschi2005}.
As a result, non-Gaussian return distributions,
sustained trends and bubbles are found,
reminiscent of real markets.
The Wealth Game uses a wealth-based payoff scheme,
and produces behaviors resembling
those of arbitrageurs and trendsetters,
and markets with positive sums~\cite{Yeung2007}.
The Wealth Game payoff scheme enables the agents
to have a strong history dependence
when applied to Hang Seng index data.

The agents in the original version of MG use a \textit{step} payoff function
\cite{Challet1997,Savit1999,Manuca2000}, meaning that the payoffs received by the winning group are the same,
irrespective of the \textit{winning margin} (the difference between the majority and the minority groups).
Latter versions of MG use a \textit{linear} payoff function
\cite{Challet2000,Marsili2000,Heimel2001,Coolen2001,Coolen2005}, in which the payoffs increase with the winning
margin. Other payoff functions yield the same macroscopic behaviors in their dependence of the population
variance on the complexity of strategies \cite{Li2000,Lee2003}. Thus, it appears that
the behavior of the population is
universal as long as the payoff function favors the minority group.

However, when one considers details beyond the population variance, one can find that the agents self-organize
in different ways for different payoff functions. For a payoff function that favors a large winning margin, the
distribution of the buyer population is double-peaked \cite{Challet1997}. This shows that the dynamics of the
population self-organizes to favor large winning margins of either buyers or sellers, because the agents have
adapted themselves to maximize their payoffs.

Another important factor is the initial preference of an agent towards the individual strategies she holds.
Considering the example of financial markets, the agents may enter the market with their own preferences of
strategies according to their individual objectives, expectations and available capital, even in the case that
they hold the same set of strategies. For example, some agents have stronger inclinations towards aggressive
strategies, and others more conservative. Hence, it is interesting to consider the effects of a diverse
distribution of strategy preferences on the system behavior. Our recent work on MG revealed that when the
diversity of strategy preferences increases, the system dynamics generally converges slowly, but the
maladaptation of the agents, as generally reflected in the fluctuations of their decisions, can be greatly
reduced \cite{Wong2004,Wong2005}.
A scaling relation between the population variance and the diversity was found,
but there are no dynamical transitions~\cite{Wong2004,Wong2005}.

Besides the diversity of preferences, the profile of the diversity distributions also influences the dynamics of
the system. For example, the Gaussian distribution of preferences studied in \cite{Wong2004,Wong2005} is a
prototype of a continuous distribution, modeling a population with less polarized opinion. This distribution is
different from the bimodal distribution studied in
\cite{Heimel2001,Coolen2001,Garranhan2000,Sherrington2002,Moro2004}, which is more appropriate to model a
population with polarized opinion. Both cases share many common statistical features, such as the reduction of
fluctuations on increasing diversity. However, since adaptation is a dynamical process, one would expect that
the bimodal case may have a much more erratic temporal behavior than the Gaussian case. This constitutes one of
the purposes of our study.

In this paper, we consider the attractor behavior of an adaptive population using linear and quadratic payoff
functions, with the distribution of initial preferences being either Gaussian or bimodal. While the statistical
behaviors in these cases are quite similar, their attractor behaviors are different, revealing the different
dynamics by which a population adapts to its specific environment. A number of findings in this paper illustrate
this point.  For example, when the step payoff function is replaced by the linear payoff function, the scaling
relation between population variance and diversity is replaced by a dynamical transition between vanishing and
non-vanishing step sizes. For low signal dimensions, the dynamical transitions take place in the form of
cascades for different signal dimensions. Even when the cascades are blurred at higher signal dimensions, it is
possible that there is a crossover from anisotropic to isotropic motion in the phase space. When the Gaussian
preference distribution is replaced by the bimodal distribution, the dynamics becomes more bursty, and the phase
space motion becomes more jumpy. Going from linear to quadratic payoffs, we find that the basins of attractor
for vanishing and non-vanishing step sizes coexist. All these rich behaviors demonstrate the flexibility of an
adaptive population for self-organizing to states in which agents maximize their payoffs and is, hence,
important in modeling of economics and distributed control.

The paper is organized as follows. After formulating MG in Section II, we consider the cases of linear and
quadratic payoffs in Sections III and IV respectively, followed by a conclusion in section V. Detailed
derivatives are presented in Appendices A to C.

\section{MINORITY GAME}

The Minority Game model consists a population of \textit{N} agents competing for limited resources, \textit{N}
being odd \cite{Challet1997}. Each agent makes a decision 1 or 0 at each time step, and the minority group wins.
For economic markets, the decisions 1 and 0 correspond to buying and selling respectively, so that the buyers
can win by belonging to the minority group, which pushes the price down, and vice versa. For typical control
tasks, such as the distribution of shared resources, the decisions 1 and 0 may represent two alternative
resources so that fewer agents utilizing a resource implies more abundance. The decisions of each agent are
responses to the environment of the game, described by \textit{signal} $\mu^{*}(t)$ at time \textit{t}, where
$\mu^{*}(t)=0,...,D-1$. These responses are prescribed by \textit{strategies}, which are binary functions
mapping the \textit{D} signals to decisions 1 or 0. In this paper, we consider both \textit{endogenous} signals
and \textit{exogenous} signals. In the endogenous case, the signals are the \textit{history} of the winning bits
in the most recent \textit{m} steps. Thus, the strategies have an input dimension of $D=2^{m}$. In the exogenous
case, the signals are randomly generated from the \textit{D} possible choices at each time step. The parameter
$\alpha\equiv\textit{D}/N$ is referred to as the \textit{complexity}.

Before the game starts, each agent randomly picks \textit{s}
strategies. Out of her \textit{s} strategies, each agent makes
decisions according to the most successful one at each step. The
success of a strategy is measured by its cumulative payoff, as
explained below.

Let $\xi_{a}^{\mu}=\pm1$ when the decisions of strategy \textit{a}
are 1 or 0, responding to signal \textit{$\mu$}. Let $a^{*}(i,t)$ be
the strategy adopted by agent \textit{i} at time \textit{t}. Then,
$$A(t)\equiv\frac{1}{N}\sum_{i}\xi_{a^{*}(i,t)}^{\mu^{*}(t)}$$ is the excess
demand of the game at time \textit{t}. The payoff received by
strategy $a$ is then $-\xi_{a}^{\mu^{*}(t)}\varphi(\sqrt{N}A(t))$,
where $\varphi$ is the payoff function. The step and linear payoffs
are $\varphi(\chi)=\mathrm{sgn}\chi$ and $\chi$, respectively.
Let $\Omega_{a}(t)$ be the cumulative payoff of strategy \textit{a}
at time \textit{t}. We will consider both \textit{online} and
\textit{batch} updates of the payoffs. The online updating dynamics
is described by
\begin{eqnarray}
\Omega_{a}(t+1)=\Omega_{a}(t)-\xi_{a}^{\mu^{*}(t)}\varphi(\sqrt{N}A(t))\label{online},
\end{eqnarray}
whereas the batch updating dynamics is given by
\begin{eqnarray}
\Omega_{a}(t+1)=\Omega_{a}(t)-\sum_{\mu}\xi_{a}^{\mu}\varphi(\sqrt{N}A^{\mu}(t))\label{batch},
\end{eqnarray}
wherein $A^{\mu}(t)$ is a $D$-dimensional vector
$$A^{\mu}(t)\equiv\frac{1}{N}\sum_{i}\xi_{a^{*}(i,t)}^{\mu}.$$

The \textit{diversity} of initial preferences of strategies is introduced
by initializing the cumulative payoffs $\Omega_{ia}(t)$
of strategy \textit{a} (\textit{a}$=2,...,s$) of agent \textit{i}
at time $t=0$ to random biases $\omega_{ia}$,
with $\Omega_{ia}(0)=0$ for $a=1$~\cite{Wong2004,Wong2005}.
Thus, the preferences influence the score of each strategy
for the rest of the game.
In this paper, we will consider both the Gaussian and delta distributions of
the initial preferences of the strategies. In the Gaussian case, the preference distribution has a mean 0 and
variance \textit{R},
\begin{eqnarray}
P(\omega_{ia})=\frac{e^{-\omega_{ia}^{2}/2R}}{\sqrt{2\pi\textit{R}}}\label{gaussdis}.
\end{eqnarray}

The ratio $\rho\equiv\textit{R}/N$ is referred to as the
\textit{diversity}. In the delta function case, $\omega_{ia}=\sqrt{\rho\textit{N}}$, as considered in
\cite{Heimel2001}. Since the system behavior is invariant with respect to random permutations of the strategies,
the introduction of the delta function is equivalent to a bimodal preference distribution
\begin{eqnarray}
P(\omega_{ia})=\delta(\omega_{ia}-\sqrt{\rho\textit{N}})/2+\delta(\omega_{ia}+\sqrt{\rho\textit{N}})/2,
\label{bimodal}
\end{eqnarray}
with mean 0 and standard deviation $\sqrt{\rho\textit{N}}$. As we shall see, the two preferences distributions
have different effects on the dynamics of the game.

To monitor the mutual adaptive behavior of the population, we
measure the variance $\sigma^{2}/N$
 of the population making decision 1, as defined by
\begin{eqnarray}
\frac{\sigma^{2}}{N}\equiv\frac{N}{4}\langle[A^{\mu^{*}(t)}(t)-\langle\mathrm{A}^{\mu^{*}(t)}(t)\rangle]^{2}\rangle\label{sigma2},
\end{eqnarray}
 where the average is taken over time when the system reaches
 steady state and over the random distribution of strategies
 and biases.

\section{MG WITH LINEAR PAYOFFS}
\subsection{THE GAUSSIAN DISTRIBUTION OF STRATEGIES' INITIAL PREFERENCES}

\begin{figure}
\centerline{\hspace{-0.5cm}\epsfig{figure=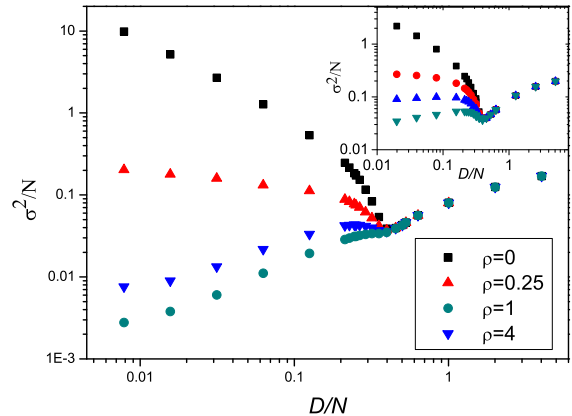, width=1.02\linewidth}} \vspace{-0.8cm}
\caption{\label{Vvsalpha}$\sigma^2/N$ versus $\alpha$ with linear payoffs for $\rho=0,0.25,1$, and 4, for
$N=251, s=2$, and 1000 samples. Inset: The corresponding plot for step payoffs.}
\end{figure}
\subsubsection{The onset of instability.}
We first consider the case of online dynamics with a Gaussian
distribution of initial preferences. As Fig.~\ref{Vvsalpha} shows,
the dependence of the variance $\sigma^{2}/N$ on the complexity
$\alpha$ for linear payoffs is very similar to that for step payoffs
\cite{Wong2004,Wong2005}. For $\alpha$ above a universal critical
value $\alpha_c$ $(\approx 0.3)$, the variance drops when $\alpha$
is reduced. The effect of introducing the diversity is also similar
to that for step payoffs; namely, the variance remains unaffected
when $\alpha>\alpha_{c}$, but decreases significantly with the
diversity when $\alpha<\alpha_{c}$.

\begin{figure} \vspace{-0.8cm}
\centerline{\hspace{-0.5cm}\epsfig{figure=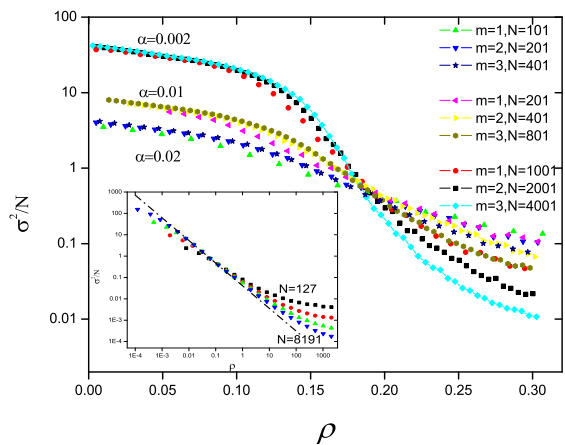, width=1.02\linewidth}} \vspace{-0.8cm}
\caption{\label{VvsP}$\sigma^2/N$ versus $\rho$ with linear payoffs of $\alpha=0.002, 0.01$, and 0.02 for
different $m$ and $N$. Inset: $\sigma^2/N$ versus $\rho$ with step payoffs $m=1$, and $N=127, 511, 2047$, and
8191. The dashed-dotted line is the scaling prediction. For both payoffs, $s=2$ and the number of samples is
1000. }
\end{figure}

However, there are differences when one goes beyond this general
trend. As Fig.~\ref{VvsP} shows, the variance curves at different
values of $\alpha$ cross at $\rho=\rho_c\approx0.16$, indicating the
existence of a continuous phase transition at $\rho_c$ from a phase
of vanishing variance at large $\rho$ to a phase of finite variance
at small $\rho$. This behavior is very different from that for step
payoffs, where the variance scales as $\rho^{-1}$ and there are no
dynamical transitions (Fig.~\ref{VvsP} inset).

\begin{figure}
\centerline{\epsfig{figure=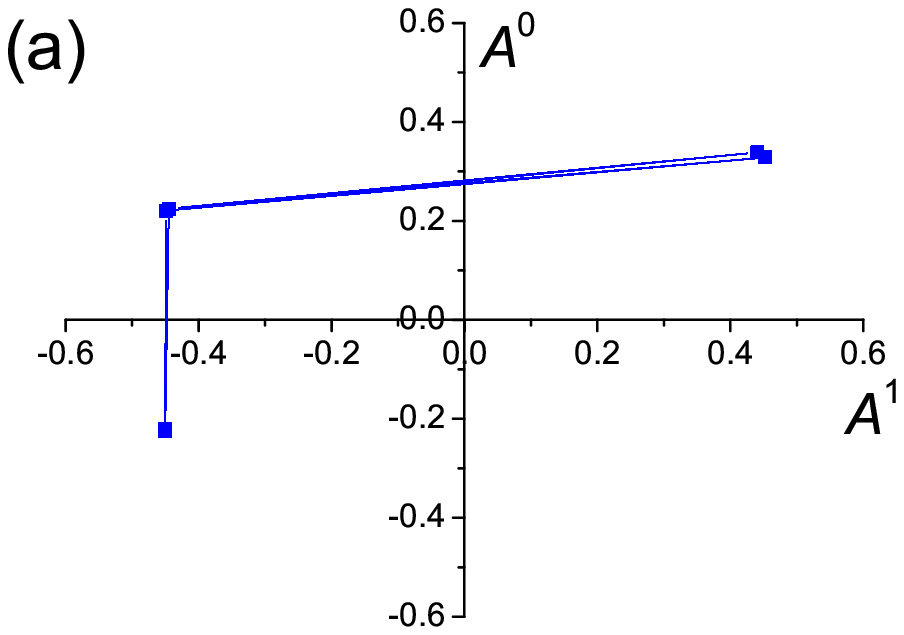, width=1\linewidth}}
\vspace{-0.3cm}
\centerline{\epsfig{figure=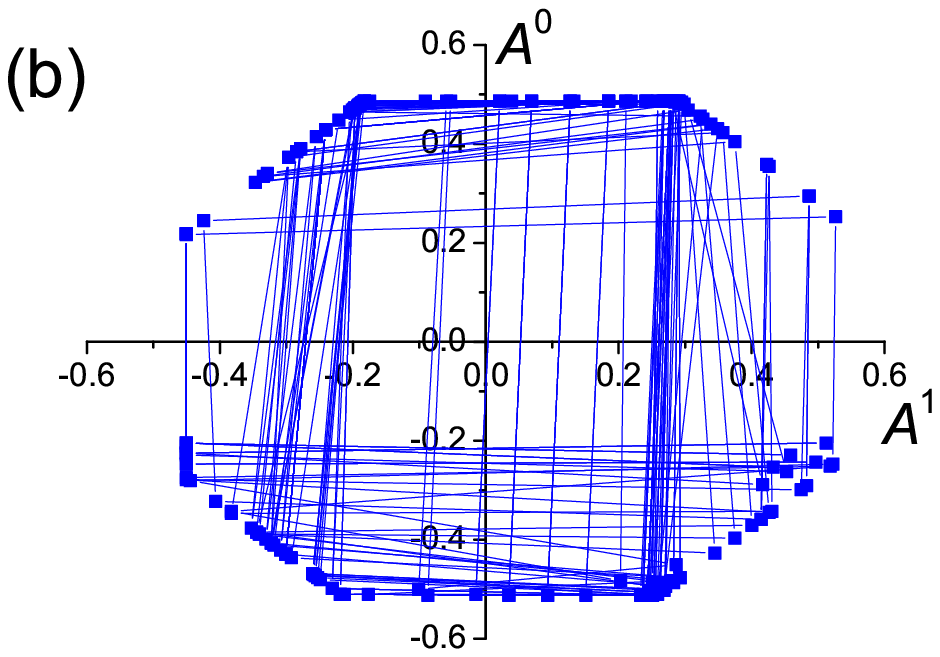, width=1\linewidth}}
\vspace{-0.3cm}
\caption{\label{layoutG}Attractor dynamics for $m=1$ in
(a) endogenous dynamics, and (b) exogenous dynamics with
online update. $\rho=0.01$}
\end{figure}

The picture is confirmed by analyzing the dynamics of the game for small $m$. The dynamics can be conveniently
described by introducing the $D$-dimensional vector $A^{\mu}(t)$ which is defined in Eq.~(\ref{batch}). While
only one of the $D$ signals corresponds to the historical signal $\mu^{*}(t)$ of the game, the augmentation to
$D$ components is necessary to describe the attractor structure of the game dynamics. Fig.~\ref{layoutG}(a)
illustrates the attractor structure in this phase space for the visualizable case of $m=1$ with online update
and endogenous signals. The dynamics proceeds in the direction that tends to reduce the magnitude of the
components of $A^{\mu}(t)$ \cite{Challet2000}. However, the components of $A^{\mu}(t)$ overshoot, resulting in
periodic attractors of period $2D$. For $m=1$, the attractor is described by the sequence $\mu^{*}(t)=0,1,1,0$,
and takes the L-shape as shown in Fig.~\ref{layoutG}(a) \cite{Wong2005}. Note that the displacements in the two
directions may not have the same amplitude. This is also true for online update and exogenous signals as
Fig.~\ref{layoutG}(b) shows, although it does not have periodic attractors like those in Fig.~\ref{layoutG}(a).

Following steps similar to those in \cite{Wong2005}, we find that for $m$ not too large and for convergence
within time steps much less than $\sqrt{R}$,
\begin{eqnarray}
A^{\mu}(t+1)=A^{\mu}(t)-\sqrt{\frac{2}{\mathrm{\pi}R}}\varphi(\sqrt{N}A^{\mu}(t))\delta_{\mu\mu^{*}(t)}.\label{step}
\end{eqnarray}
For step payoffs, Eq.~(\ref{step}) converges to an attractor confined in a $D$-dimensional hypercube of size
$\sqrt{2/{\mathrm{\pi}R}}$, irrespective of the value of $R$. On the other hand, for linear payoffs,
$A^{\mu}(t+1)$ becomes a linear function of $A^{\mu}(t)$ with a slope of $1-\sqrt{2/\pi\rho}$. Hence, for
$\rho>\rho_{c}=1/{2\pi}\sim0.16$, the \textit{step sizes}
$$\Delta\textit{A}^{\mu}(t)\equiv\mid\textit{A}^{\mu}(t+1)-A^{\mu}(t)\mid$$ converge to zero. In particular for
$1/{2\pi}<\rho<2/\pi$, the motion in the phase space converges with oscillations, whereas for $\rho>2/\pi$, the
motion converges without oscillations. On the other hand, for $\rho<\rho_{c}$, steps of vanishing sizes become
unstable, resulting in a continuous dynamical transition at $\rho_{c}$.

The phase transition at $\rho_{c}$ illustrates how the agents adapt to the changing environment when the
diversity changes. In general, at high diversity, only a small fraction of agents switches their strategies at
each time step. This gentle movement results in, at most, small winning margins, as revealed in the example of
step payoffs \cite{Wong2004,Wong2005}. In contrast, the winning margins at low diversity can be larger. Since we
are considering payoffs that are linear functions of the winning margin, the agents are adapted to pay more
attention to decisions that result in larger profits, leading to stronger responses at low diversity and
vanishing responses at high diversity.

\subsubsection{The time scales of convergence.}
The onset of the instability when $\rho$ decreases is accompanied by the separation of two convergence times.
The first time scale is the {\it state convergence time}, which is defined as the number of time steps needed to
reach the attractor (for $m=1$, this is the sequence 0,1,1,0)~\footnote{For $m=1$, we discard the cases which
do not converge to the attractor, which weighs about 10 percent of all samples. For the cases that converge to
the attractor, the convergence time is measured as the first time it reaches the attractor, regardless of
whether it will jump out subsequently.}.

\begin{figure}
\centerline{\hspace{-0.5cm}\epsfig{figure=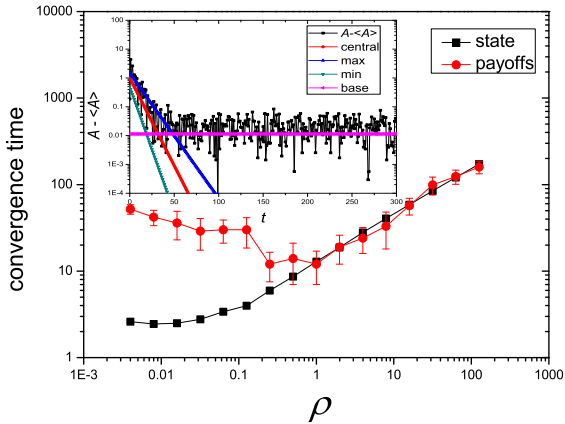, width=1\linewidth}} \vspace{-0.8cm}
\caption{\label{convergtime} The dependence of the state and population convergence time on the diversity for
online update, endogenous dynamics, linear payoffs and Gaussian preference distribution. Inset: the relation
between $A-\langle\textit{A}\rangle$ and \textit{t} from the transient to the steady state for $\rho=7.943$.
Parameters: $m=1, s=2$, and $N=255$, averaged over 1000 samples.}
\end{figure}

The second time scale is the {\it population convergence time} which is defined for online update and endogenous
dynamics as follows. Note that the excess demand $A(t)$ may not arrive at the steady-state time series even
after the attractor states $-\mathrm{sgn}A(t)$ become steady. To measure the convergence time scale for $A(t)$,
the periodic trend of $A(t)$ in the attractor must be subtracted. This is done by measuring the sample average
of $\overline{\textit{A}_{i}}=\langle\textit{A}(t)\rangle_{mod(t,2D)=i}$ for $i=0,...,2D-1$ and monitoring the
time dependence of $\textit{A}(t)-\overline{\textit{A}}_{mod(t,2D)}$. As shown in the inset of
Fig.~\ref{convergtime}, this difference converges exponentially to a baseline, and the inverse of the slope of
the exponential convergence yields the population convergence time. As shown in Fig.~\ref{convergtime}, the
state convergence time increases smoothly with diversity. On the other hand, the population convergence time is
distinct from the state convergence time in the low diversity region. The turning point is between $\rho=0.13$
and 0.25, indicating a relation with the dynamical transition of step sizes at $\rho_{c}$.

\subsubsection{Cascades of transitions.}
\begin{figure}
\centerline{\hspace{-0.5cm}\epsfig{figure=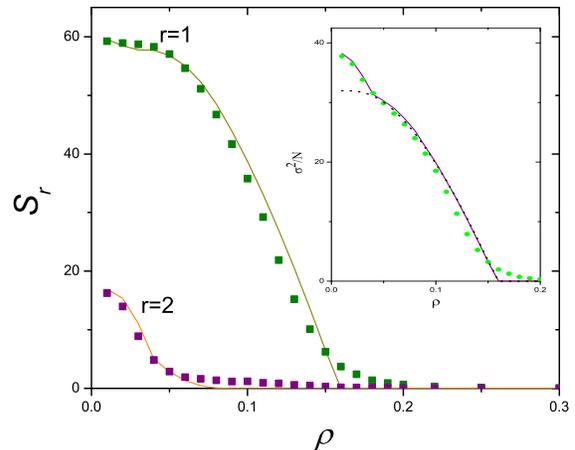, width=1\linewidth}} \vspace{-0.8cm}
\caption{\label{cascade2}$S_{r}$ versus $\rho$ for online update, endogenous dynamics, linear payoffs and
Gaussian preference distribution. Inset: $\sigma^{2}/N$ versus $\rho$. The symbols are the simulation, the
dotted line the theory with one bifurcation, and the solid lines the theory with two bifurcations. $N=1001$,
$m=1$, $s=2$, and 1000 samples.}
\end{figure}

\begin{figure}
\vspace{-0cm}
\centerline{\hspace{-0.5cm}\epsfig{figure=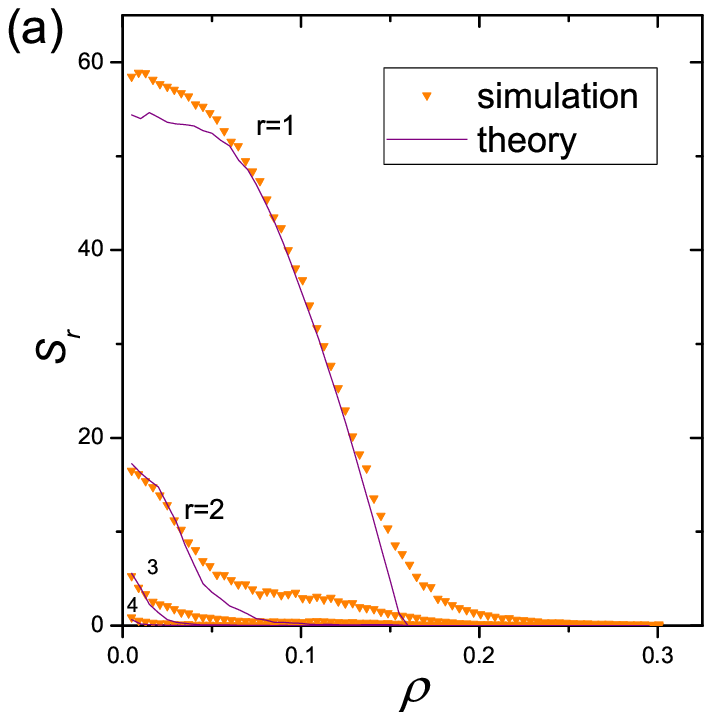,width=1\linewidth,
height=0.8\linewidth}}
\vspace{-0.3cm}
\centerline{\hspace{-0.5cm}\epsfig{figure=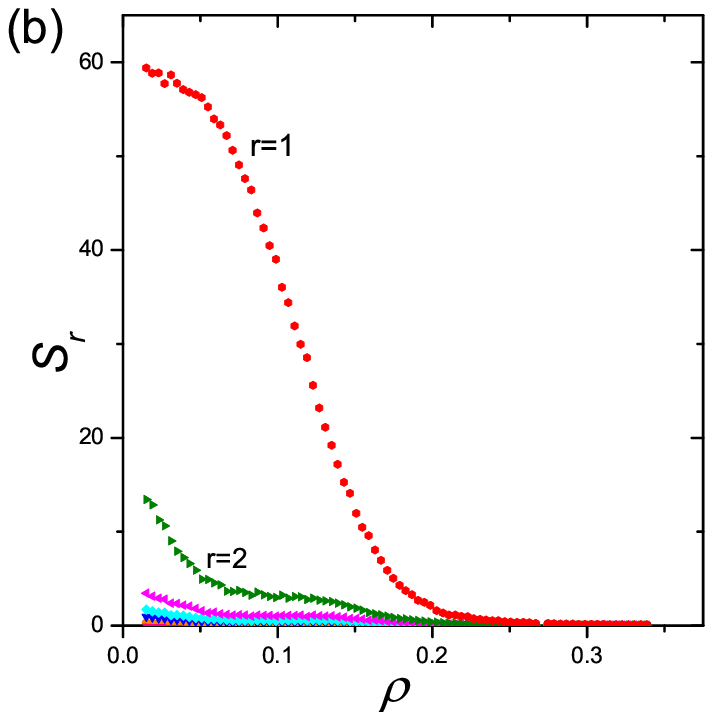,width=1\linewidth,
height=0.8\linewidth}}
\vspace{-0.3cm}
\caption{\label{cascade48}$S_{r}$ versus $\rho$ for online update, endogenous dynamics, linear payoffs and
Gaussian preference distribution(a) $m=2$, (b) $m=3$. In both cases, $N=1001$, $s=2$, and 1000 samples.}
\end{figure}

 However, when $\rho<\rho_{c}$, the step sizes for each of the
$D$ signals may not be equal. To see this, we monitor the variance
for each of the $D$ signals and rank them. The \textit{r}th maximum
variance is then given by
\begin{eqnarray}
S_{r}=\mathrm{large}_{\mu}\left(\frac{N}{4}[\langle(\textit{A}^{\mu})^2\rangle|_{\mu=\mu^{*}(t)}-(\langle\textit{A}^{\mu}\rangle|_{\mu=\mu^{*}(t)})^2],r\right),
\end{eqnarray}
where $\mathrm{large}_{\mu}(f(\mu),r)$ is the \textit{r}th largest function $f(\mu)$ for $\mu=0,...,D-1$. As
Figs.~\ref{cascade2}-\ref{cascade48} show, the step sizes for the $D$ signals do not bifurcate simultaneously at
$\rho=\rho_{c}$, Rather, only the first maximum bifurcates from zero when $\rho$ falls below $\rho_{c}$, while
the step sizes for the remaining $D-1$ signals remain small. When the diversity further decreases to around
0.05, the second maximum becomes unstable as well, and a further bifurcation takes place. For $m\geq2$, there
are further bifurcations of the third or higher order maxima, resulting in a \textit{cascade} of dynamical
transitions when the diversity decreases.

This cascade of transitions is confirmed by analysis. For $m=1$, we can generalize Eq.~(\ref{step}) to
convergence times of the order $\sqrt{R}$. Assuming, without loss of generality, that $\textit{A}^{1}$
bifurcates while $\textit{A}^{0}$ remains small. We find that the variance of the buyer population, as derived
in Appendix A, is
\begin{eqnarray}
\frac{\sigma^{2}}{N}=\frac{N}{32}(\Delta\textit{A}^{1})^{2},
\quad\Delta\textit{A}^{1}=\mathrm{erf}\left(\frac{\Delta\textit{A}^{1}}{\sqrt{8\rho}}\right)\label{32},
\end{eqnarray}
where $\Delta\textit{A}^{1}$ is the step size responding to signal 1. As the inset in Fig.~\ref{cascade2} shows,
the analytical and the simulated results agree down to $\rho \sim 0.05$. However, when the diversity decreases
further, this simple analysis implies that the variance will saturate to a constant $N/32$, but simulated
results are clearly higher. This discrepancy is due to a further bifurcation of the minimum step size. This can
be analyzed by considering the effect of a perturbation $\delta\textit{A}^{0}(t)$ in the direction of
$\textit{A}^{0}$. As derived in Appendix B, the accumulated perturbation becomes
\begin{eqnarray}
\delta\textit{A}^{0}(t+4)=\left[1-\frac{1}{\sqrt{2\pi\rho}}(1+e^{-\frac{(\Delta\textit{A}^{1})^{2}}{8\rho}})\right]^{2}\delta\textit{A}^{0}(t)\label{stability}.
\end{eqnarray}
At $\rho=0.0459$, where $\Delta\textit{A}^{1} = 0.9775$, the coefficient on the right-hand side of
Eq.~(\ref{stability}) reaches the value 1, and $\delta\textit{A}^{0}(t)$ diverges on further reduction of
$\rho$. Numerical iterations of the analytical equations for $\textit{A}^{\mu}(t)$, averaged over samples of
different initial conditions, yield the theoretical curves in Fig.~\ref{cascade2} and the inset, agreeing very
well with the simulated results. Similarly, the agreement between analytical and simulated results are
satisfactory for $m=2$, as shown in Fig.~\ref{cascade48}(a).

For $m=3$, as shown in Fig.~\ref{cascade48}(b), the bifurcation of $\textit{S}_{1}$, remains distinctive from
those of other directions, and the picture of at least one cascade is valid. Furthermore, we observe that in the
low diversity limit for $m=1,2,3,$ the maximum step size approaches the same value of $N/16$ following the
arguments in Appendix A. We also want to mention that results show that although the attractor structure of
online update with exogenous signals is different from that with endogenous signals,
the dependence of $\textit{S}_{r}$ and $\sigma^2/N$ on $\rho$ are totally the same for both signals.



This shows that the cascades are a general feature of the adaptive dynamics of the agents. When the diversity
decreases, the agents find it more profitable to induce large winning margins responding to certain, but not all
signals. At sufficiently low diversity, large winning margins become possible responses to all, signals,
resulting in the cascades of transitions. However, for large signal dimensions at higher \textit{m}, the
interference of the responses to different signals will blur the transitions.

\subsubsection{Batch update}
\begin{figure}
\centerline{\hspace{-0.5cm}\epsfig{figure=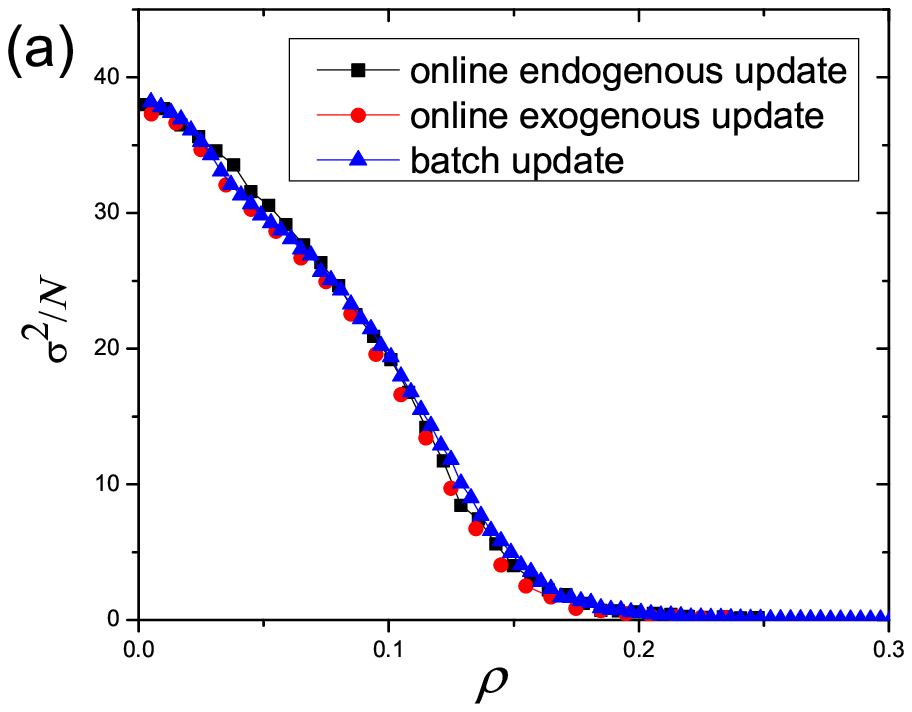,width=1\linewidth}}
\vspace{-0.3cm}
\centerline{\hspace{-0.5cm}\epsfig{figure=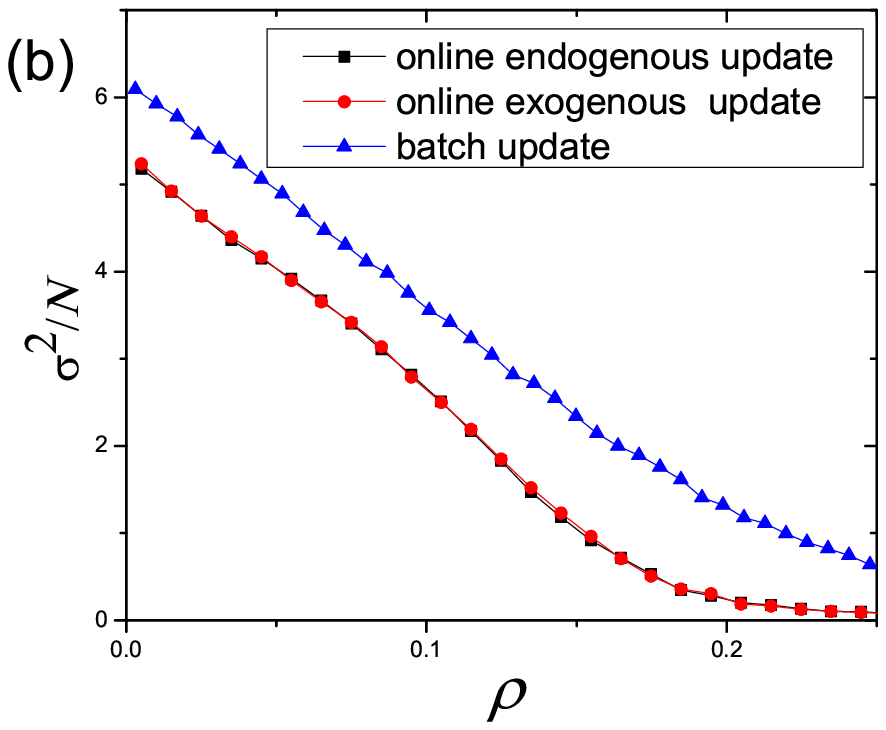,width=1\linewidth}}
\vspace{-0.3cm}
\caption{\label{batchcom}$\sigma^2/N$ versus $\rho$ for linear payoffs and Gaussian preference distribution. (a)
$m=1$; (b) $m=4$. In both cases, $N=1001, s=2$, and 1000 samples.}
\end{figure}

We close this section briefly mentioning the results of batch update. We found that the variance of the buyer
population with batch update is larger than that with online update (those of endogenous and exogenous signals
update are the same) and the difference increase as $\alpha$ increases, as shown in Fig.~\ref{batchcom}. We also
found that the dynamical transition to non-vanishing step sizes in the case of online update was replaced by a
gradual crossover around $\rho\sim0.2$, and the cascades of transitions in different directions are replaced by
a simultaneous crossover around $\rho\sim0.2$.

\subsection{THE BIMODAL DISTRIBUTION OF STRATEGIES' PREFERENCES}

In this section we consider how the dynamics depends on
different distributions of agents' initial preferences of strategies.
In the previous section we have studied
the case of Gaussian distributions of preferences,
which model populations with less polarized opinions.
In this section, we replace the Gaussian distribution
with a bimodal distribution described by Eq.~(\ref{bimodal}),
which is more appropriate to model a population with polarized opinions.
As we shall see, the latter shares many common statistical features
with the former, such as
the cascades of dynamical transitions (subsection 1)
and the change in the isotropy of motion (subsection 2)
when the diversity changes.
On the other hand,
the latter exhibits more erratic temporal behavior,
such as the changing fraction of time spent on locations
of high and low volatility in the attractor (subsection 3)
and the bursty dynamics (subsection 4).
The theoretical dynamical equations
for both batch and online updates are derived in Appendix C.

\subsubsection{Cascades of dynamical transitions}

\begin{figure}
\centerline{\hspace{-0.5cm}\epsfig{figure=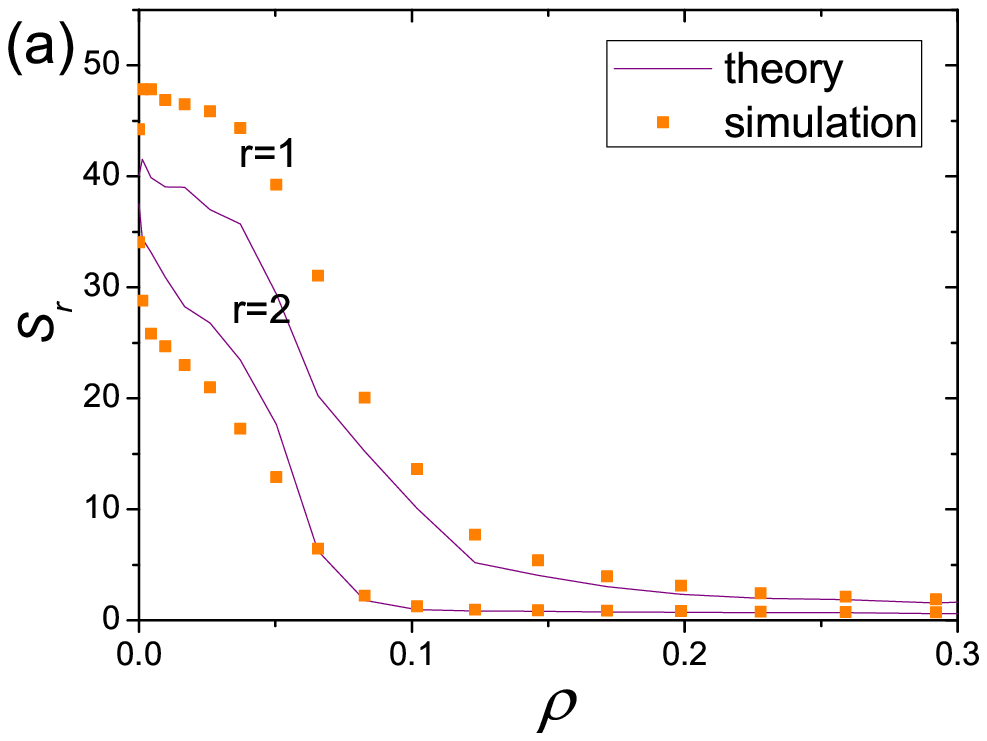,width=1\linewidth}}
\vspace{-0.3cm}
\centerline{\hspace{-0.5cm}\epsfig{figure=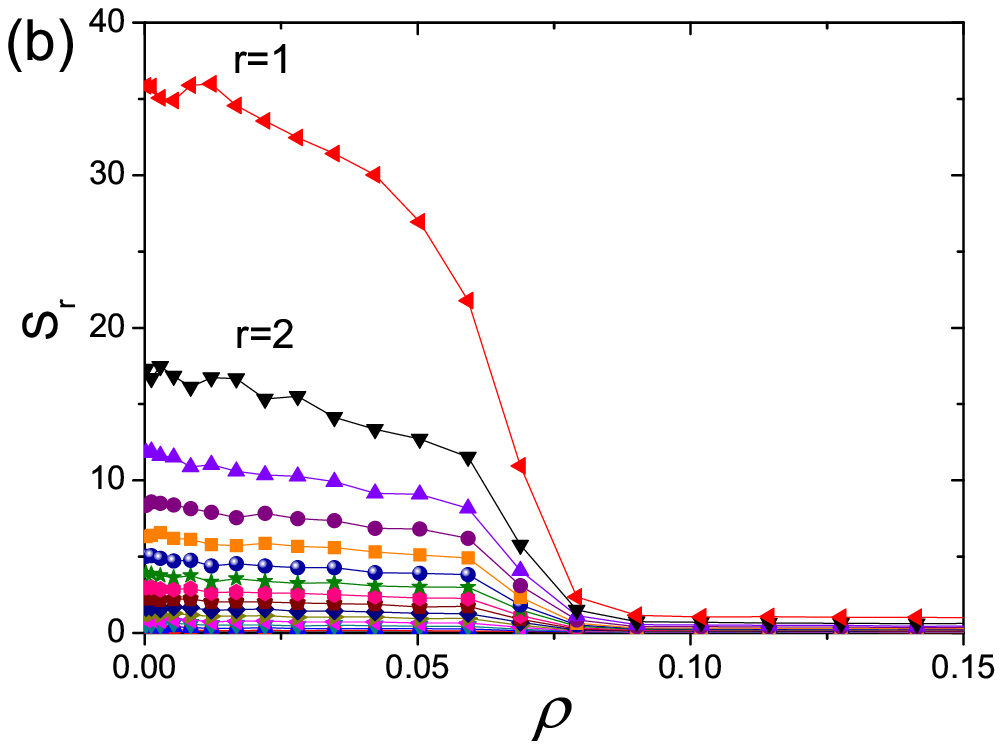,width=1\linewidth}}
\vspace{-0.3cm}
\caption{\label{Graphbatch}$S_{r}$ versus $\rho$ for batch update, linear payoffs and bimodal preference
distribution (a) $m=1$ and (b) $m=4$. In both cases, $N=1001$, $s=2$, 1000 samples, and batch updates.}
\end{figure}

We found that for batch updates, Fig.~\ref{Graphbatch}(a) shows that the cascade of dynamical transitions is
present for small signal dimensions but disappears as signal dimensions increase. For large signal dimensions,
step sizes for different signal dimensions bifurcate simultaneously at $\rho\sim0.09$. As shown in
Fig.~\ref{Graphbatch}(b), there is a large jump of $S_{1}$, indicating that it is a discontinuous transition,
while for the Gaussian case, it is a continuous transition. This discontinuous transition was found previously
in \cite{Heimel2001}, and the transition point is around $\rho_{c}\sim0.06$ with the magnitude of the jump
scaling as $\alpha^{-\frac{1}{2}}$. Results are similar for online update.

\subsubsection{Isotropy of motion}

For batch update, we have devised new global parameters to describe whether the motion in the phase space is
isotropic at each update. For $D=2$, the two parameters are
$$U_{1}\equiv\langle\langle[(\Delta\textit{A}^{1})^{2}-(\Delta\textit{A}^{0})^{2}]^{2}\rangle_{t}\rangle_{\textrm{sample}}$$
and
$$V_{1}\equiv\langle\langle[2\Delta\textit{A}^{1}\Delta\textit{A}^{0}]^{2}\rangle_{t}\rangle_{\textrm{sample}}, $$
they measure the displacements along the axial or diagonal directions in the phase space of $A^{1}$ and $A^{0}$.

\begin{figure}[t]
\leftline{\epsfig{figure=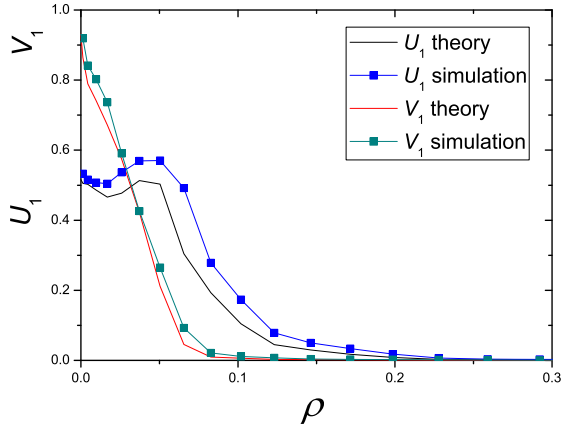,width=\linewidth}} \vspace{-0.5cm} \caption{\label{Graphdiahv}$U_{1}$ and
$V_{1}$ versus $\rho$ for batch update, linear payoffs and bimodal preference distribution, $m=1,s=2,N=1001$,
and 1000 samples.}
\end{figure}

We can see from Fig.~\ref{Graphdiahv} that as diversity decreases, the axial parameter grows from nearly zero to
nonzero values at around 0.2, while the diagonal parameter remains small. This shows that at each time step, the
system responds to only one signal. When $\rho$ decreases to about 0.06, the diagonal parameter becomes nonzero
too, showing that the system responds to more than one signal at each time step, and the motion becomes more
isotropic. This cascade behavior is also well supported by our theory displayed on the same figure.

\begin{figure*}
\centering
\leftline{\epsfig{figure=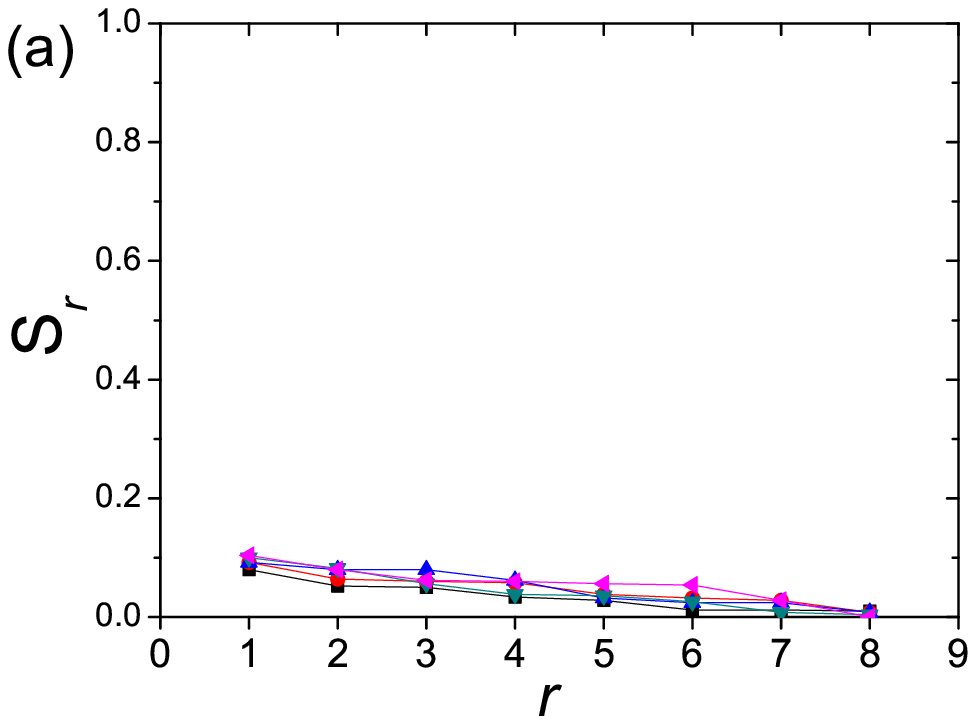,width=0.47\linewidth}
\leftline{\epsfig{figure=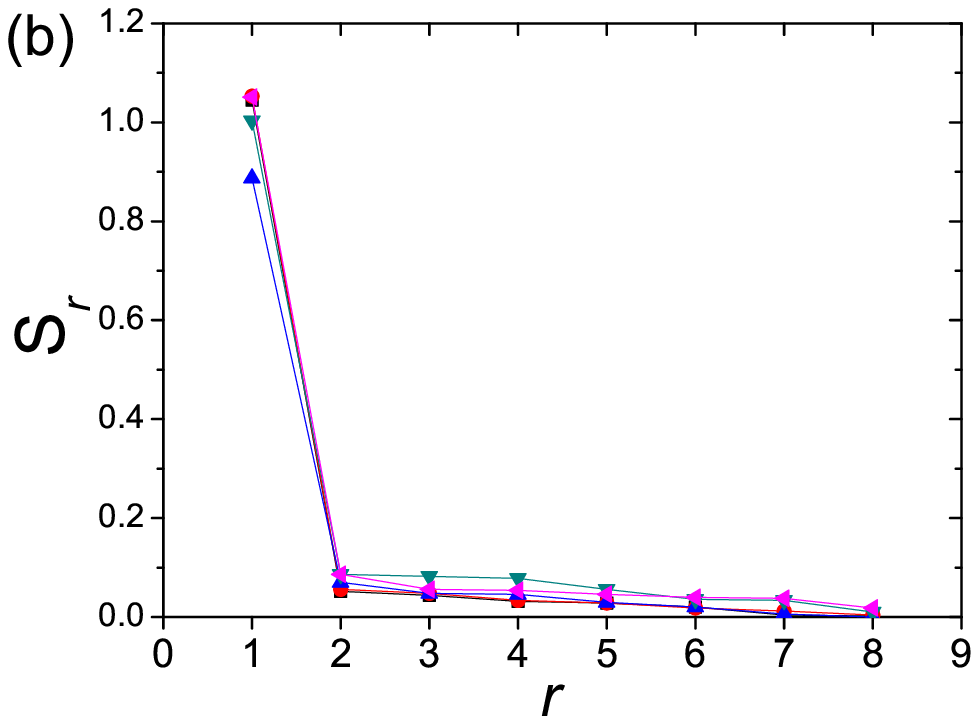,width=0.47\linewidth}}}
\leftline{\epsfig{figure=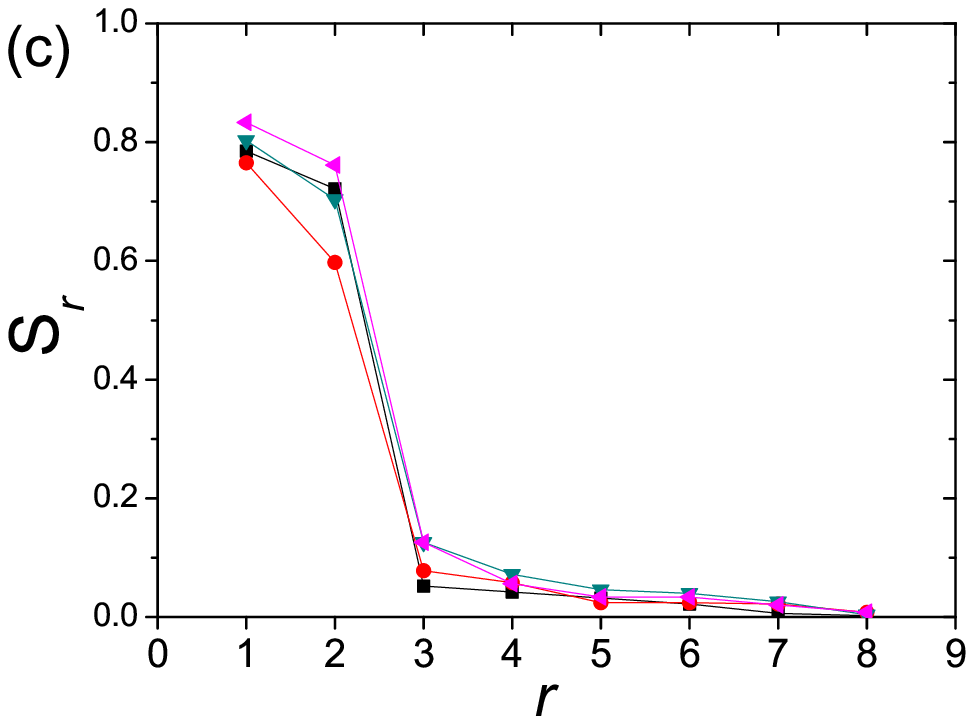,width=0.47\linewidth}
\leftline{\epsfig{figure=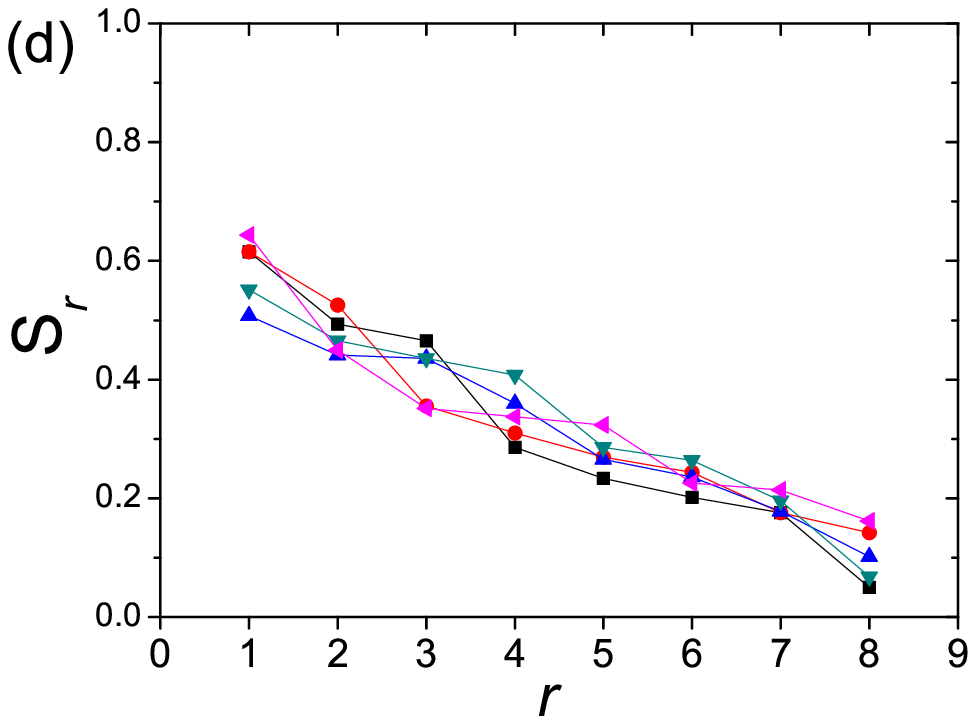,width=0.47\linewidth}}}

\caption{\label{Graphrankatt}(a) to (d) Ranked step sizes for four classes of behavior with increasing isotropy,
in batch update, linear payoffs and bimodal preference distribution. $m=3, s=2, N=1001, \rho=0.04$, and 200
samples.}
\end{figure*}

For higher signal dimensions, the isotropy of motions in the attractors can be described by ranking the
\textit{D} components of randomly picked step sizes. The ranked components of different samples, as shown in
Fig.~\ref{Graphrankatt}, exhibit several classes of behavior with increasing isotropy. Samples in
Fig.~\ref{Graphrankatt}(a) correspond to small steps in all dimensions. Those in Fig.~\ref{Graphrankatt}(b)
correspond to steps with one large component and $D-1$ small components. Steps in Figs.~\ref{Graphrankatt}(c)
and (d) have, respectively, two and more than two large components.


\begin{figure}[t]
\leftline{\epsfig{figure=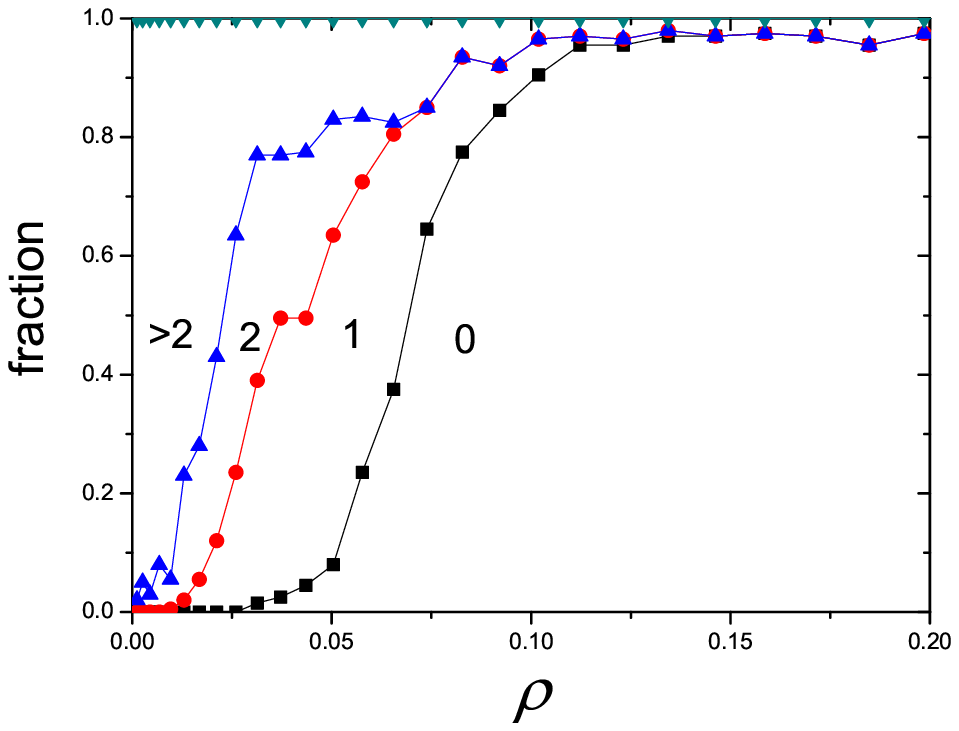,width=\linewidth}} \vspace{-0cm} \caption{\label{Graphfrac}The fraction of
the four classes of step sizes as a function of $\rho$, in batch update, linear payoffs and bimodal preference
distribution. Other parameters: same as those in Fig.~\ref{Graphrankatt}.}
\end{figure}

As shown in Fig.~\ref{Graphfrac}, attractors with small step sizes are dominant in the region of large $\rho$.
Step sizes increase when  $\rho$ decreases, but the non-vanishing components do not increases isotropically.
Rather, steps with one or two non-vanishing components become significant when $\rho\sim0.2$, and become more
isotropic on further reduction of  $\rho$.

This sequential onset of isotropy is consistent with the cascade of dynamical transitions for the Gaussian
online case when the diversity decreases. However, when $m$ increases further, interferences between different
signal dimensions will blur the anisotropic attractors.

\subsubsection{Attractor structure}
\begin{figure}[t]
\centerline{\epsfig{figure=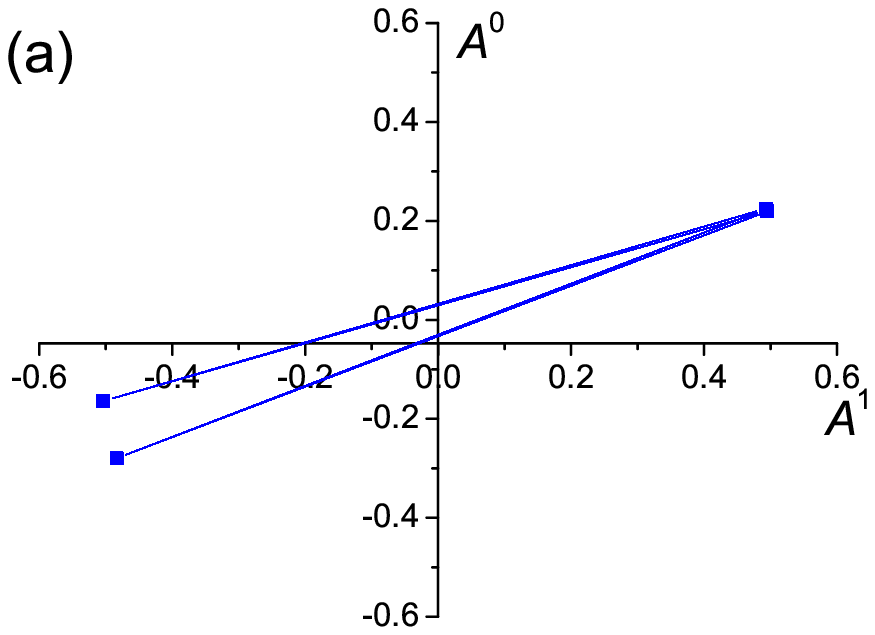,width=\linewidth}} \vspace{-0.5cm}
\centerline{\epsfig{figure=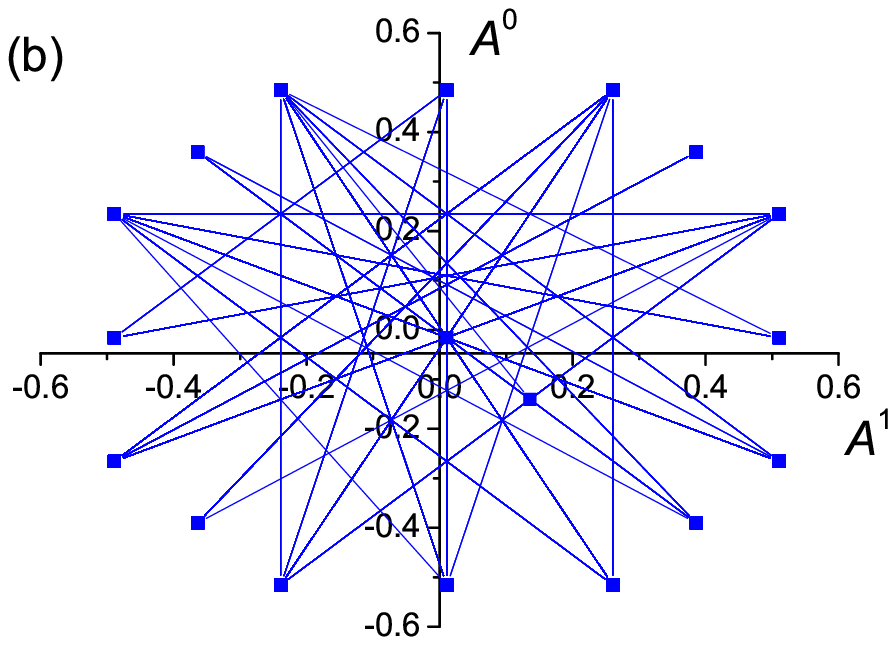,width=\linewidth}} \vspace{-0.5cm}
\centerline{\epsfig{figure=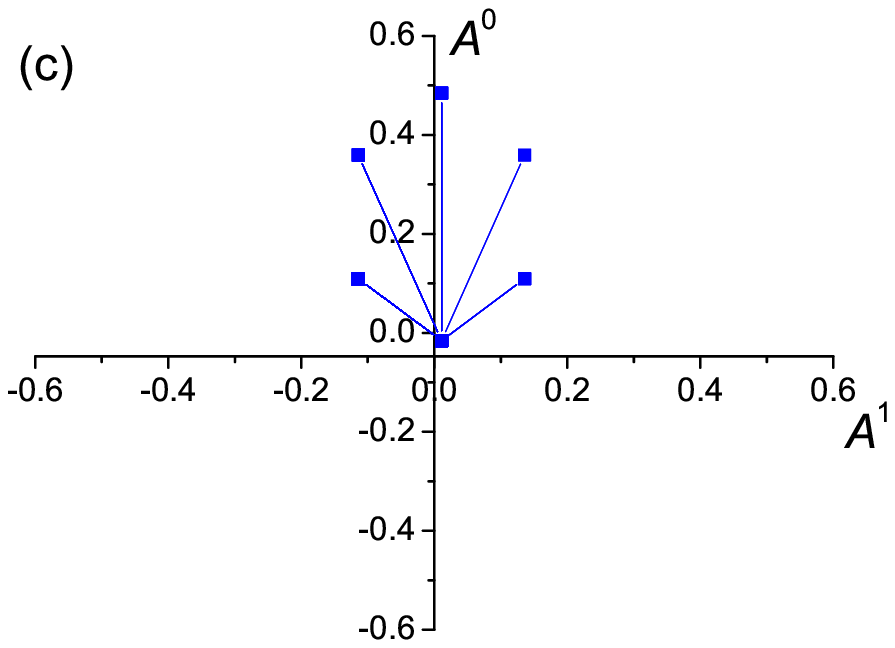,width=\linewidth}} \vspace{-0.5cm}
\caption{\label{Graphbimatt}The attractors of $m=1$ in the phase space for batch update and linear payoffs:(a)
Gaussian preference distribution, $\rho=0.01$; (b)bimodal distribution for small $\rho$ ($\rho=0.004$) and
(c)large $\rho$ ($\rho=0.04$) respectively. Both for batch update.}
\end{figure}

The structures of the attractors also change with $\rho$. For convenience, we take the example of $m=1$, whose
phase space can be plotted in two dimensions. From Fig.~\ref{Graphbimatt}, we can see that the attractor
structure with bimodal distribution is totally different from that with Gaussian distribution. For systems with
Gaussian distribution and batch update, the attractor has only two clusters of fixed points, between which
systems oscillate (Fig.~\ref{Graphbimatt}(a)), whereas for systems with bimodal distribution, the attractor
visits many points located around an octagon in the small $\rho$ phase ($\rho<\rho_{c}$), as shown in
Fig.~\ref{Graphbimatt}(b). The system jumps among these points and occasionally stays at the origin. However in
the large $\rho$ phase ($\rho>\rho_{c}$), the system will spend more time around the origin with occasional
jumping out and back.

\subsubsection{Bursty dynamics}
\begin{figure}[t]
\leftline{\epsfig{figure=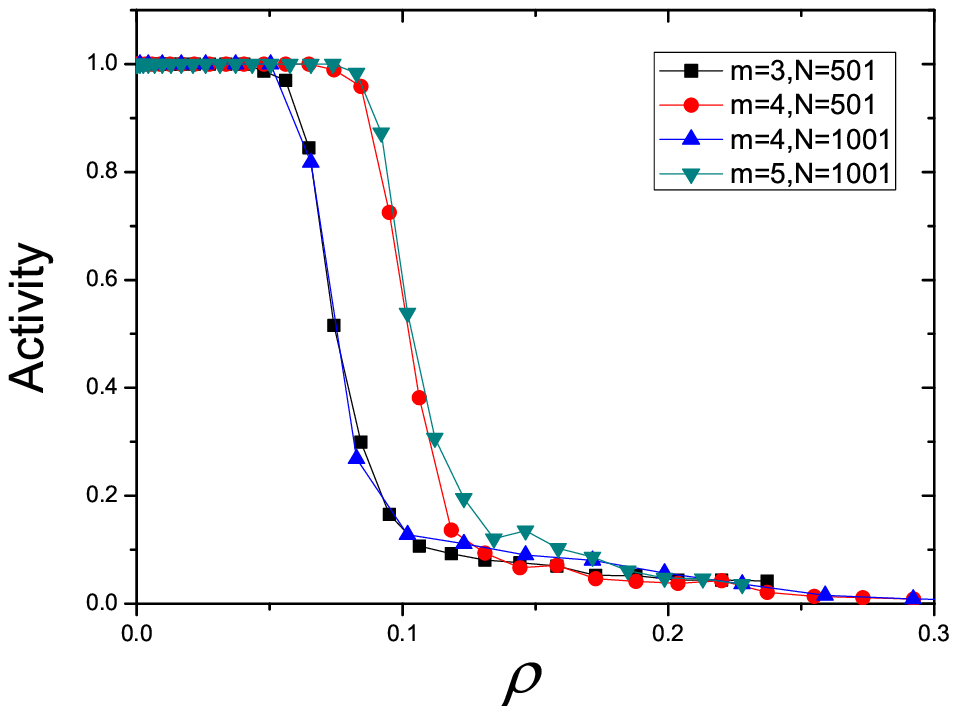,width=\linewidth}} \vspace{-0.5cm} \caption{\label{Graphlifetime}The
activity as a function of diversity for bimodal distribution of preferences, batch update and linear payoffs.}
\end{figure}

Unlike the case of Gaussian distributions, the transition from vanishing to non-vanishing step sizes on
decreasing diversity takes place in a bursty manner. For a given diversity, we define the \textit{activity} as
the fraction of time that the system stays away from the origin in the phase space. Fig.~\ref{Graphlifetime}
shows that the activity is low at high diversities, but increases to a high value when $\rho$ is reduced below a
critical value $\rho_{c}$. This critical value $\rho_{c}$ depends on $\alpha$. Estimating $\rho$ as the point
with an activity of 0.5 in Fig.~\ref{Graphlifetime}, $\rho_{c}\sim0.07$ and 0.1 for $\alpha=0.016$ and 0.032
respectively. We anticipate that $\rho_{c}$ approaches the value of 0.06 in the limit vanishing $\alpha$, as
proposed by \cite{Heimel2001}.

\begin{figure}
\centerline{\epsfig{figure=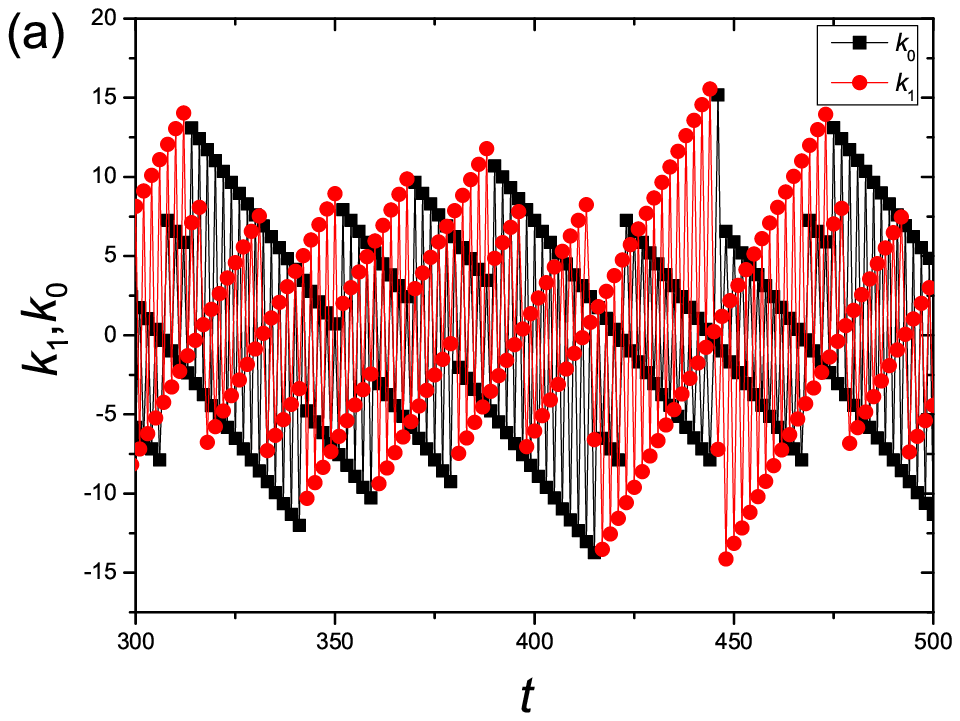,width=\linewidth}} \vspace{-0.5cm}
\centerline{\epsfig{figure=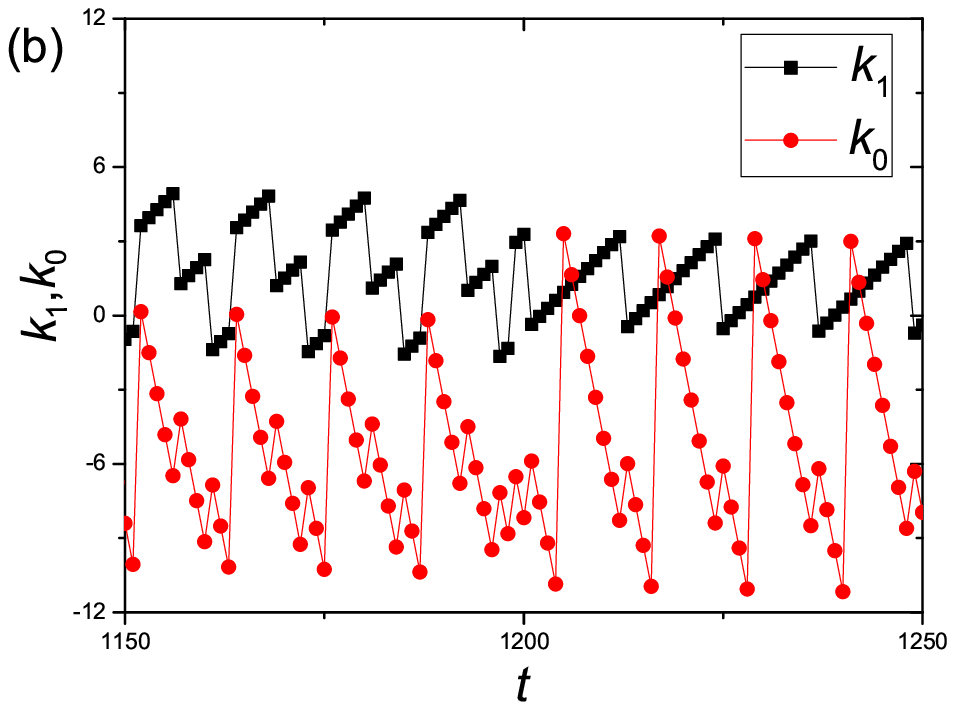,width=\linewidth}} \vspace{-0.5cm}
\caption{\label{Graphkq}The payoff components $k_{0}$ and $k_{1}$ as a function of time \textit{t} of a typical
sample with bimodal distributions of preferences, batch update and linear payoffs. Parameters are $m=1$, $s=2$,
and $N=1001$, for (a) $\rho=0.01$; (b) $\rho=0.4$.}
\end{figure}

Looking deeper into the dynamics, we can find bursty behavior by introducing the payoff components $k_{\mu}(t)$,
which are related to the accumulative payoffs by
\begin{eqnarray}
\Omega_{a}(t)=\sum_{\mu}k_{\mu}(t)\xi_{a}^{\mu}.\label{payoffcomp}
\end{eqnarray}
In other words, $k_{\mu}(t)$ is the total payoff of decision 1 of strategy \textit{a} for signal $\mu$ during
the history of the game up to time $t$. For $m=1$, the payoff components $k_{0}$ and $k_{1}$ also have different
behaviors at low and high diversities. As shown in Fig.~\ref{Graphkq}(a) for low diversity, both $k_{0}$ and
$k_{1}$ oscillate around 0 with large step sizes, resulting in the phase with non-vanishing variance. In
contrast, Fig.~\ref{Graphkq}(b) shows that for high diversities, the payoffs accumulate with small step sizes,
building up gradually to high values, and then return to low values with a huge step size. This bursty process
resembles many natural phenomena in which energy is stored gradually and released suddenly, such as earthquakes
and volcano eruptions.

\section{MG WITH QUADRATIC PAYOFF FUNCTIONs}
\begin{figure}
\centerline{\epsfig{figure=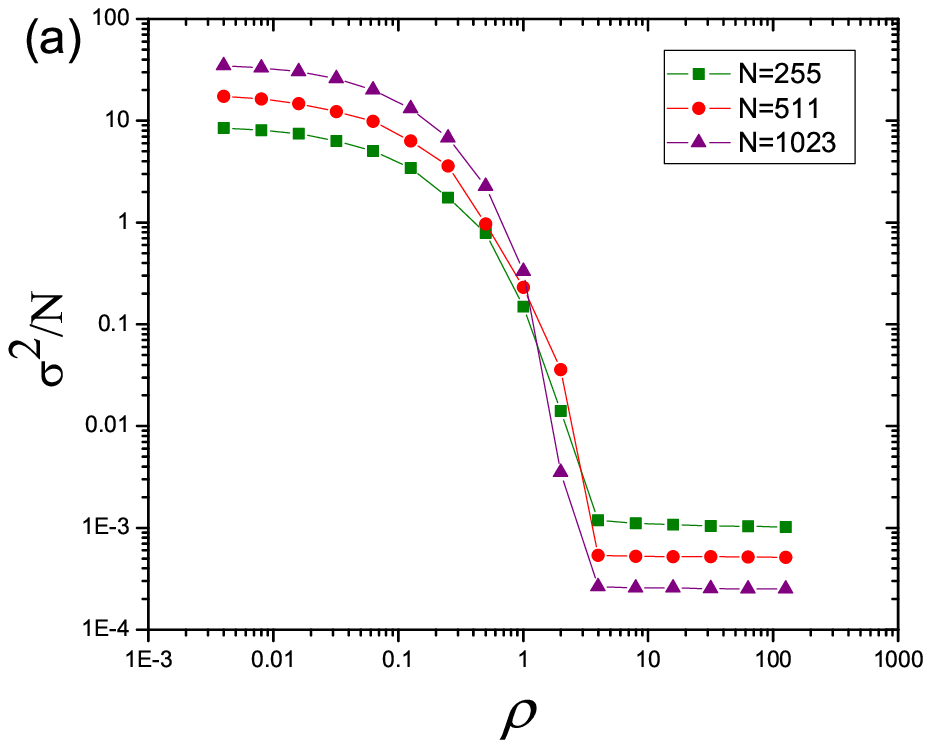,width=\linewidth}} \vspace{-0.5cm}
\centerline{\epsfig{figure=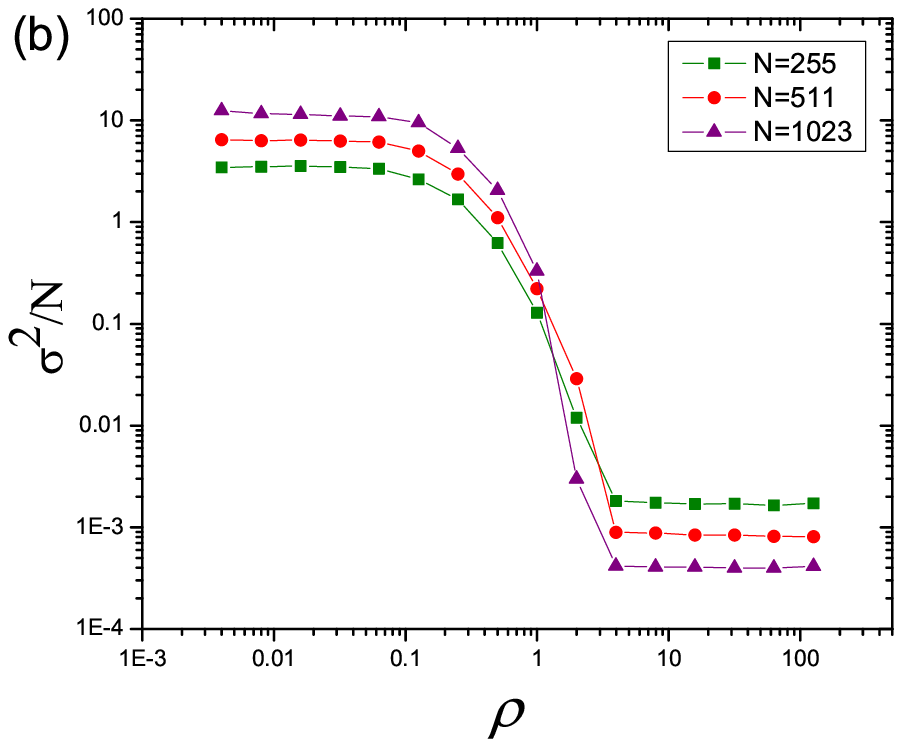,width=\linewidth}} \vspace{-0.5cm}
\caption{\label{quadraticV}The dependence of variance on the diversity for online update, quadratic payoffs,
endogenous dynamics and Gaussian preference distribution at (a) $s=2$, $m=1$ and averaged over 1000 samples. (b)
$s =2, m =2$ and averaged over 1000 samples.}
\end{figure}

To see how the behavior depends on the payoff functions, we change the payoff to the form of
$\varphi(x)=x^{2}sgn(x)$ where $x=\sqrt{N}A(t)$. Figure~\ref{quadraticV} shows the general trend that a greater
diversity gives a smaller variance of attendance and this effect is especially sensitive in the intermediate
diversity region. It does not show any scaling behavior like in the step payoff model, nor any phase transition
in the linear payoff model. Instead, it shows that a larger population always have a greater drop in the
variance and a lower minimum variance.

\begin{figure}
\leftline{\epsfig{figure=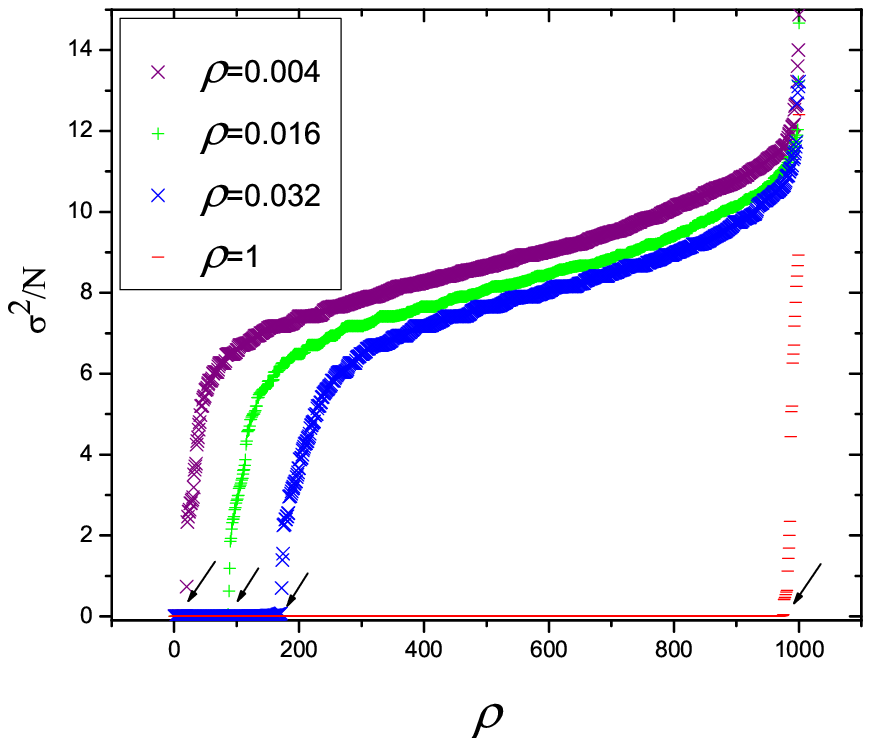,width=\linewidth}} \vspace{-0.5cm} \caption{\label{distriorder}The variance of
1000 samples arranged in ascending order for $\rho=0.004,0.008,0.032$, and 1, for online update, quadratic
payoffs, endogenous dynamics and Gaussian preference distribution. In all cases, $N=255$, $s=2$, and $m=1$.}
\end{figure}

Besides, we are also interested to investigate how the initial
position of the game affects its variance. First, 1000 samples were
simulated for each of the several different diversities. Then the
variance for the 1000 samples was arranged in ascending order and
listed graphically in Fig.~\ref{distriorder}.

We can see that more and more samples have a relatively small
variance as the diversity becomes larger. For each curve in
Fig.~\ref{distriorder}, the arrangement of the variance shows a gap
in the variance, as pointed by an arrow. To facilitate the
explanation, we define samples that are left to the gap to be "small
variance", while those that are right to the gap to be "large
variance".

\begin{figure*}
\centering
\leftline{\epsfig{figure=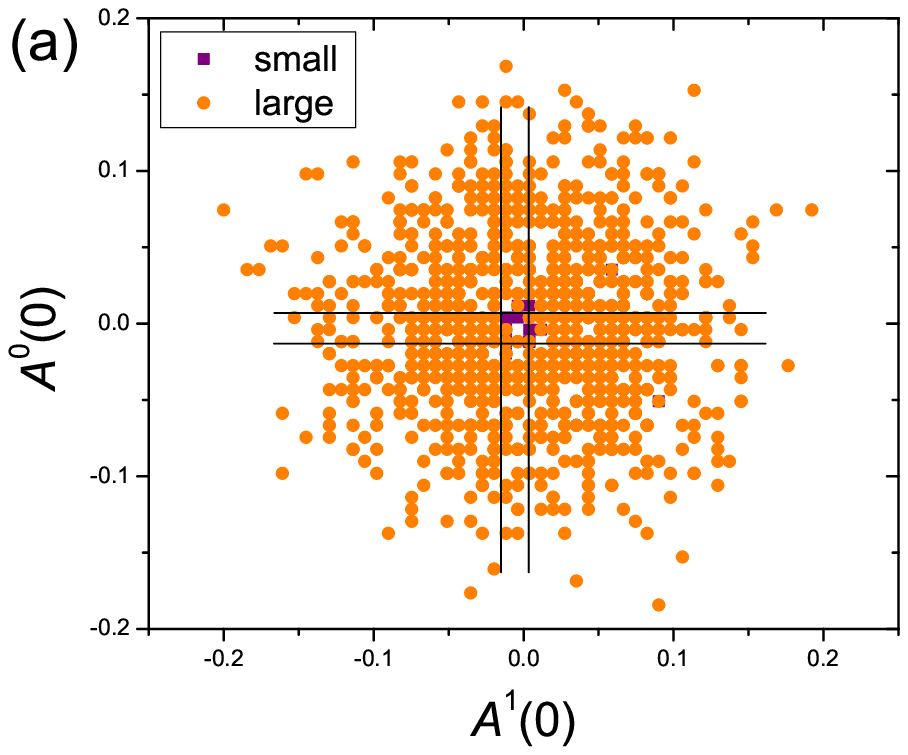,width=0.47\linewidth}
\leftline{\epsfig{figure=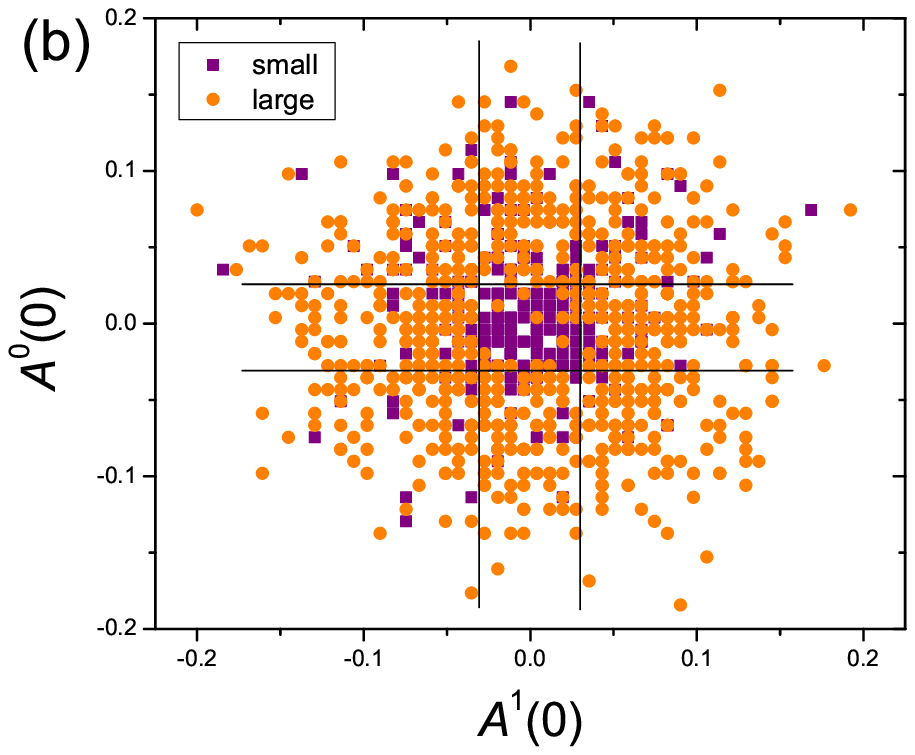,width=0.47\linewidth}}}
\leftline{\epsfig{figure=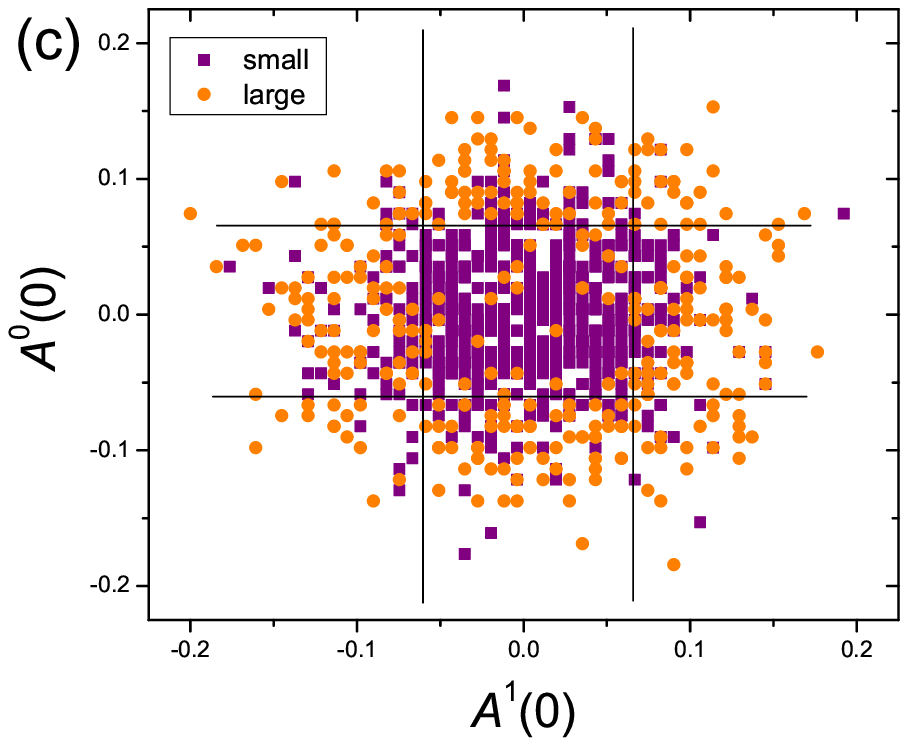,width=0.47\linewidth}
\leftline{\epsfig{figure=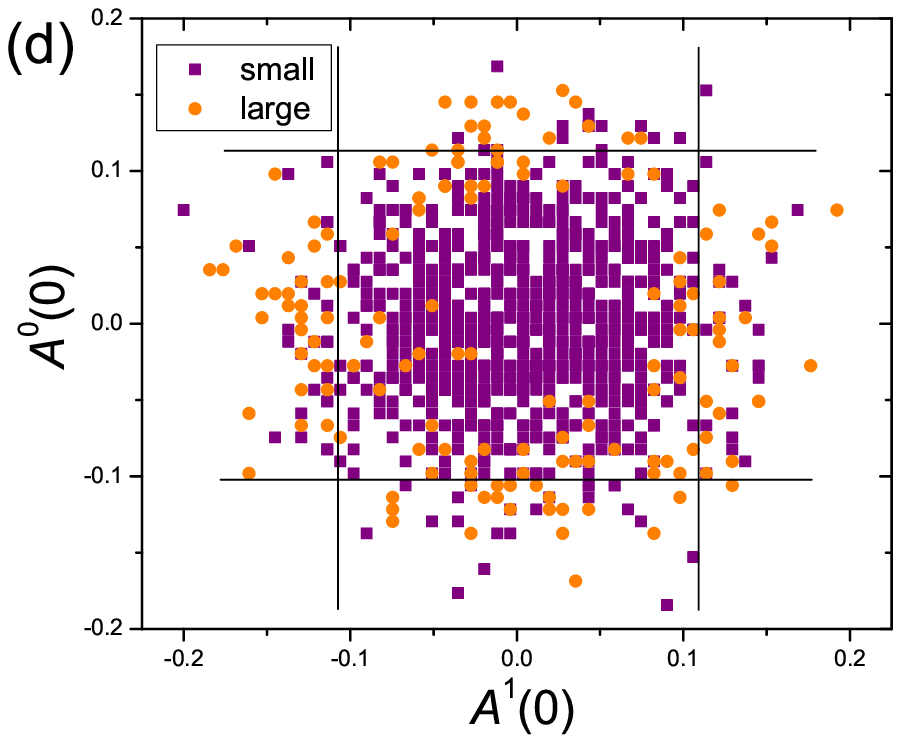,width=0.47\linewidth}}}

\caption{\label{diswithinitial}(color online) The distribution of the variance related to its initial position
for online update, quadratic payoffs, endogenous dynamics and Gaussian preference distribution, $N=255, s=2,
m=1$ and (a) $\rho=0.004$. (b) $\rho=0.063$. (c) $\rho=0.251$. (d) $\rho=0.501$.}
\end{figure*}

We find that these two groups of samples are strongly correlated with the initial states of the system. From
Figs.~\ref{diswithinitial}(a) to (d), it shows a general trend that the initial positions of the small variance
samples concentrate around the origin while those of the large variance samples spread around. A basin boundary
which is indicated by the lines appears among the light (orange) and dark (pink) dots and it becomes more
recognizable at higher diversity. Also, we can notice that more samples with small variance and less samples
with large variance appear as the diversity increases.

To determine the boundary analytically, we should look into the
dynamics of the payoff function, Following the steps similar to
those in Section III, we found

\begin{eqnarray}
A^\mu(t+1)=A^\mu(t)-\sqrt{\frac{2}{\pi\mathrm{R}}}(\sqrt{N}A^\mu(t))^2\mathrm{sgn}A^\mu(t)\label{quadraticAmu}
\end{eqnarray}

The square boundary can then be determined through the intersection between the Eq.~(\ref{quadraticAmu}) and the
lines $y=-x$.

\begin{eqnarray}
-A^\mu(t)=A^\mu(t)-\sqrt{\frac{2}{\pi\mathrm{R}}}(\sqrt{N}A^\mu(t))^2
\end{eqnarray}
or
\begin{eqnarray}
\sqrt{N}A^\mu(t)=\pm\sqrt{2\pi\rho}\label{boundary}
\end{eqnarray}

If the initial position of the system is located inside this basin boundary, that is,
$|NA^{\mu}(0)|<\sqrt{2\pi\textit{R}}$, $A^{\mu}(t)$ will eventually converge to the origin, which implies that
the agents are not motivated to respond to the low payoffs. If the initial position of the system locates
outside this basin boundary, $A^{\mu}(t)$ will eventually diverge, It implies that the agents are motivated to
respond to high payoffs. Also, from Eq.~(\ref{boundary}), the boundary increases with diversity $\rho$. Thus, a
larger diversity will give a smaller size of the basin of attraction for samples with large variance.

\begin{figure}
\leftline{\epsfig{figure=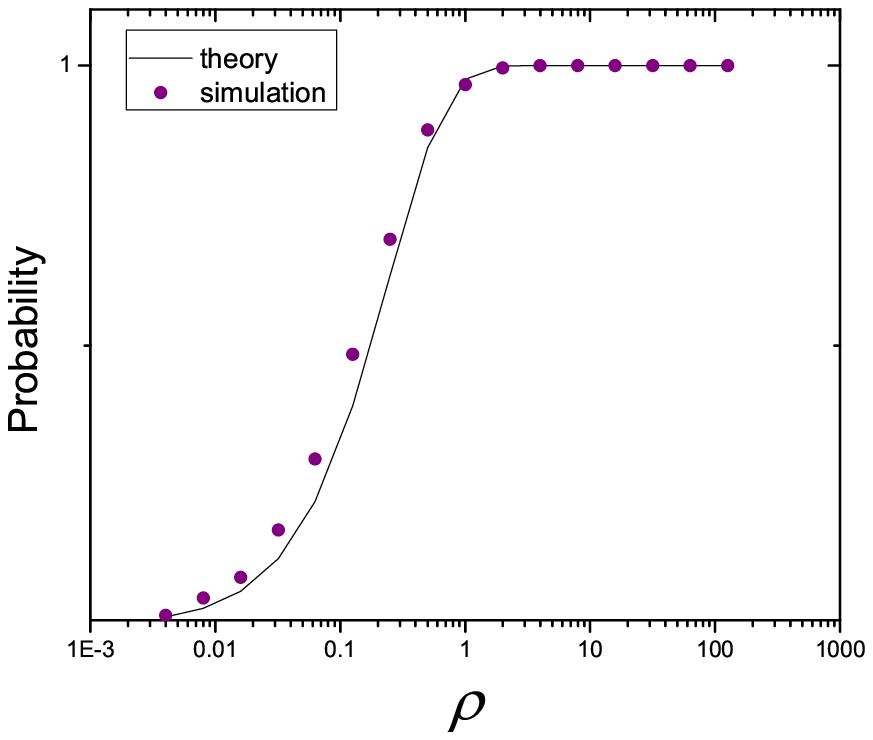,width=\linewidth}} \vspace{-0.5cm} \caption{\label{Probcomp}The comparison of
the theoretical and simulative probabilities of finding samples with small variance for online update, quadratic
payoffs, endogenous dynamics and Gaussian preference distribution, $N=255, s=2$, and $m=1$.}
\end{figure}

After locating the basin boundary, we can compute the probability of
finding attractors with small variance and large variance,
respectively. Since the probability density function of the initial
state $\sqrt{N}A^\mathrm{\mu}(0)$ is Gaussian with mean $=0$ and
variance $=1$, the probability of finding small variance attractor
is given by
\begin{eqnarray}
P_\mathrm{small}=(\mathrm{erf}\sqrt{\pi\rho})^{2}
\end{eqnarray}
and the probability of finding large variance attractor is
\begin{eqnarray}
P_\mathrm{large}=1-P_\mathrm{small}.\nonumber
\end{eqnarray}
This result is consistent with the simulation in that when the
diversity is large, the probability of finding samples with small
variance will be higher. A comparison between theory and simulation
is given in Fig.~\ref{Probcomp}, which shows a good agreement.

\section{Conclusion}

We have studied the behavior of an adaptive population by using a payoff function that increases linearly and
quadratically with the winning margin. We found in linear payoffs, a continuous dynamical transition when the
adaptation rate of the population was tuned by varying their diversity of preferences. This is in contrast with
the case of payoff functions independent of the winning margin, in which there is only a scaling relation
between the variance and the diversity, and no phase transitions are found. The dynamical transition is due to
the payoffs being enhanced by large winning margins at low diversity. Furthermore, for systems with
multi-dimensional signals feeding the strategies, we found a cascade of dynamical transitions in the responses
to different signals, with the population variance increasing at each cascade. When the cascades are blurred at
higher signal dimensions, a classification of the step vectors as done in Fig.~\ref{Graphrankatt} also reveals
the possibility of a crossover from anisotropic to isotropic motion in the phase space.

We have also studied the effects of polarization of the initial preference distribution on the attractor
behavior of the system by comparing the Gaussian and bimodal distributions. In the Gaussian case, there is a
gradual increase in step sizes when the diversity decreases, whereas in the bimodal case, there is rather a
sharp increase in the \textit{fraction of time} with non-vanishing sizes in a narrow range of diversity, as
shown in Fig.~\ref{Graphlifetime}. We also found that for bimodal distributions, the payoff components go
through processes of accumulation and bursts. These observations illustrate that the more polarized the
preferences are, the more erratic the dynamics is. There are many similarities shared among the different cases
that we have studied. For both Gaussian and bimodal distributions, there are cascades for low signal dimensions
which gradually disappear as signal dimensions increase. For online update, both endogenous and exogenous
signals result in similar macroscopic behaviors, for example, the diversity dependence of the variance of the
buyer population and the ranked step sizes.

For quadratic payoffs, a basin boundary separating two groups of samples is found. The group of vanishing step
sizes adapts to the low payoff region of the quadratic payoff function and the group of non-vanishing step sizes
adapts to the region with more rapidly increasing payoffs of the quadratic payoff function.

In summary, we have found three pathways through which fluctuations increase with decreasing diversity. For
linear payoffs with Gaussian preference distribution, fluctuations increase by increasing step sizes via
cascades of continuous dynamical transitions. For linear payoffs with bimodal distributions, fluctuations
increase is also due to the increase in the \textit{fraction of time} with non-vanishing step sizes. For
quadratic payoffs, fluctuations increase by enlarging the basin of the attractors with large fluctuations. This
illustrates the rich behavior that the population can self-organize in different environments.

It is interesting to consider the phenomenology
experienced by the agents at each side
of the described dynamical transitions and crossovers.
In this respect, we make the following observations.

1) Agents using linear payoffs
focus more of their attention
on opportunities of large winning margins,
and tend more to neglect marginal wins.
In markets with high diversity,
agents with different preferences
switch their strategies at different times,
preventing the occurrence of a high volatility.
Since the volatility is not large enough
to induce strategy switches
of such agents focusing more on large winning margins,
the market is self-organized
to a state of low volatility.
However, in markets with low diversity,
more agents switch their strategies at the same time,
and the volatility is high.
These profitable opportunities are exploited
by the agents more sensitive to large winning margins,
and the market is self-organized
to a state of high volatility.
In markets with intermediate diversity,
the agents are selectively sensitive to some,
but not all, of the signals in the market,
leading to the cascades of phase transitions.

2) For agents using quadratic payoffs,
their interest on marginal wins vanishes even faster.
Hence the market is self-organized
to a state of low volatility
when the initial volatility is low,
irrespective of the diversity of preferences of strategies.
On the other hand,
their emphasis on large winning margins
rises even faster than linear payoffs.
Hence the market is self-organized
to a state of high volatility
when the initial volatility is high,
irrespective of the diversity.
Changes in the diversity cannot preclude
the attractors of either high or low volatility.
They merely influence the sizes
of their basins of attraction.

3) Both cases of linear and quadratic payoffs are different from the case of step
payoffs~\cite{Wong2004,Wong2005}. Agents using step payoffs place an equal emphasis on all winning opportunities
irrespective of the winning margin. Consequently, the volatility does not vanish at any values of diversity, and
there are no dynamical transitions or multiple basins of attraction. Instead. a scaling relation between the
volatility and the diversity is applicable.

4) In markets with a bimodal distribution of initial preferences,
the agents have polarized opinions
about their responses to the signals.
When the diversity is high, 
it takes a large number of time steps
before the opinions of a group of agents are reversed.
This gives rise to periods of vanishing volatility,
during which the system stays
at the origin of the phase space.
When the opinions of the agent group is eventually reversed,
bursts of activities erupt.

Recent attention was drawn to the role of the payoff function on reproducing realistic market behavior such as
non-Gaussian features, the formation of sustained trends and bubbles, and intermittency \cite{de
Martino2004,Tedeschi2005}. Our study further confirms that tuning the payoff function and the preference
distribution can lead to a rich spectrum of self-organized states of the market. It would be interesting to
consider populations of agents with different individual payoff functions and study how they interact.

\appendix
\section{The equations of motion in MG with linear payoffs}
We consider the equations of motion in MG
with linear payoffs. Here, we focus on the case of a Gaussian distribution of initial preferences, and the
generalization to the case of the bimodal distribution is straightforward. Using Eq.~(\ref{online}), and
averaging over Eq.~(\ref{gaussdis}), we obtain the step size for the historical state $\mu=\mu^{*}(t)$.
\begin{eqnarray}
\langle\Delta\textit{A}^{\mu}(t)\rangle&&=\frac{1}{2^{2D-1}}\sum_{a<b}\int\frac{\textit{d}\omega}{\sqrt{2\pi\textit{R}}}e^{-\omega^{2}/2R}(\xi_{a}^{\mu}-\xi_{b}^{\mu})\nonumber\\
&&\times[\Theta(\omega+\Omega_{a}(t+1)-\Omega_{b}(t+1))\nonumber\\
&&-\Theta(\omega+\Omega_{a}(t)-\Omega_{b}(t))]\label{deltaA}.
\end{eqnarray}
Since the integral representation of the step function is given by
\begin{eqnarray}
\Theta(y)=\int_{0}^{\infty}\textit{d}x\int\frac{\textit{d}p}{2\pi}e^{ip(x-y)},
\end{eqnarray}
Eq.~(\ref{deltaA}) becomes
\begin{eqnarray}
\langle\Delta\textit{A}^{\mu}(t)\rangle&&=\frac{1}{2^{2D-1}}\sum_{a<b}\int\frac{\textit{d}\omega}{\sqrt{2\pi\textit{R}}}e^{-\omega^{2}/2R}(\xi_{a}^{\mu}-\xi_{b}^{\mu})\nonumber\\
&&\times\!\int_{0}^{\infty}\!\textit{d}x\int\frac{\textit{d}p}{2\pi}e^{ip(\omega-x)}[e^{ip(\Omega_{a}(t+1)-\Omega_{b}(t+1))}\nonumber\\
&&-e^{ip(\Omega_{a}(t)-\Omega_{b}(t))}]\label{deltaA2}.
\end{eqnarray}

Decomposing the cumulative payoffs into payoff components as defined in Eq.~(\ref{payoffcomp}),
Eq.~(\ref{deltaA2}) becomes

\begin{eqnarray}
\langle\Delta\textit{A}^{\mu}(t)\rangle&&=\frac{1}{2^{2D-1}}\sum_{a<b}\int\frac{\textit{d}\omega}{\sqrt{2\pi\texttt{R}}}e^{-\omega^{2}/2R}\nonumber\\
&&\times\int_{0}^{\infty}\textit{d}x\int\frac{\textit{d}p}{2\pi}e^{ip(\omega-x)}\prod_{\nu\neq\mu}e^{ipk_{\nu}(t)(\xi_{a}^{\nu}-\xi_{b}^{\nu})}\nonumber\\
&&\times[e^{ip(k_{\mu}-\sqrt{N}A^{\mu})(\xi_{a}^{\mu}-\xi_{b}^{\mu})}-e^{ipk_{\mu}(\xi_{a}^{\mu}-\xi_{b}^{\mu})}]\nonumber\\
&&\times(\xi_{a}^{\mu}-\xi_{b}^{\mu})\label{deltaA3}.
\end{eqnarray}

Using the identity for $\xi_{a}^{\mu},\xi_{b}^{\mu}=\pm1$ and
$$e^{i\theta(\xi_{a}^{\mu}-\xi_{b}^{\mu})}=\cos^{2}\theta+i\sin\theta\cos\theta(\xi_{a}^{\mu}-\xi_{b}^{\mu})+\sin^{2}\theta\xi_{a}^{\mu}\xi_{b}^{\mu},$$ we arrive at
\begin{eqnarray}
&&\langle\Delta\textit{A}^{\mu}(t)\rangle=\int_{0}^{\infty}\textit{d}x\int\frac{\textit{d}p}{2\pi}e^{-\frac{R}{2}p^{2}-ip\textit{x}}\times\nonumber\\
&&\prod_{\nu\neq\mu}\cos^{2}pk_{\nu}[i\sin2p(k_{\mu}-\sqrt{N}A^{\mu})\!-\!i\sin2pk_{\mu}]\label{deltaAmu}.
\end{eqnarray}
where we have used the identity $\sum_{a}\xi_{a}^{\mu}=0$.

Similarly, for the non-historical states $\nu\neq\mu^{*}(t)$, we
have

\begin{eqnarray}
&&\langle\Delta\textit{A}^{\nu}(t)\rangle=\int_{0}^{\infty}\textit{d}x\int\frac{\textit{d}p}{2\pi}e^{-\frac{R}{2}p^{2}-\textit{i}p\textit{x}}\nonumber\\
&&\times\prod_{\lambda\neq\mu\nu}\cos^{2}pk_{\lambda}2i\sin\textit{p}k_{\nu}\cos\textit{p}k_{\nu}\nonumber\\
&&\times[\cos^2p(k_{\mu}-\sqrt{N}A^{\mu})-\cos^2pk_{\mu}]\label{deltaAmuba}.
\end{eqnarray}

For $m=1$, there are only two states. Let $\mu$ and $\overline{\mu}$
be the historical and non-historical states respectively.

After evaluating the integrals, we obtain
\begin{eqnarray}
\langle\Delta\textit{A}^{\mu}(t)\rangle&&=\frac{1}{8}\Biggl[\mathrm{erf}\frac{2k_{\mu}(t+1)+2k_{\overline{\mu}}(t)}{\sqrt{2R}}\nonumber\\
&&+\mathrm{erf}\frac{2k_{\mu}(t+1)-2k_{\overline{\mu}}(t)}{\sqrt{2R}}+2\mathrm{erf}\frac{2k_{\mu}(t+1)}{\sqrt{2R}}\nonumber\\
&&-\mathrm{erf}\frac{2k_{\mu}(t)+2k_{\overline{\mu}}(t)}{\sqrt{2R}}-\mathrm{erf}\frac{2k_{\mu}(t)-2k_{\overline{\mu}}(t)}{\sqrt{2R}}\nonumber\\
&&-2\mathrm{erf}\frac{2k_{\mu}(t)}{\sqrt{2R}}\Biggr]\label{deltaA13}.
\end{eqnarray}

\begin{eqnarray}
\langle\Delta\textit{A}^{\overline{\mu}}(t)\rangle&&=\frac{1}{8}\Biggl[\mathrm{erf}\frac{2k_{\mu}(t+1)+2k_{\overline{\mu}}(t)}{\sqrt{2R}}\nonumber\\
&&-\mathrm{erf}\frac{2k_{\mu}(t+1)-2k_{\overline{\mu}}(t)}{\sqrt{2R}}-\mathrm{erf}\frac{2k_{\mu}(t)+2k_{\overline{\mu}}(t)}{\sqrt{2R}}\nonumber\\
&&+\mathrm{erf}\frac{2k_{\mu}(t)-2k_{\overline{\mu}}(t)}{\sqrt{2R}}\Biggr]\label{deltaA14},
\end{eqnarray}

where $k_{\mu}(t+1)=k_{\mu}(t)-\sqrt{N}A^{\mu}(t)$.

\begin{figure}
\leftline{\hspace{-0.25cm}\epsfig{figure=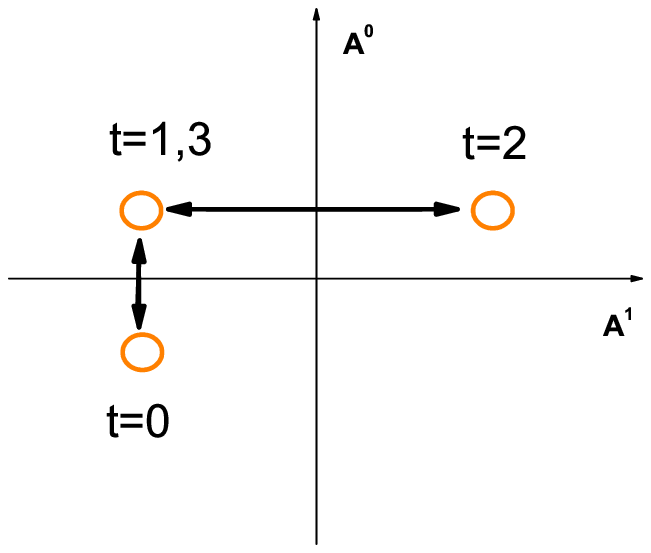,
width=0.53\linewidth}
\hspace{-0.3cm}\leftline{\epsfig{figure=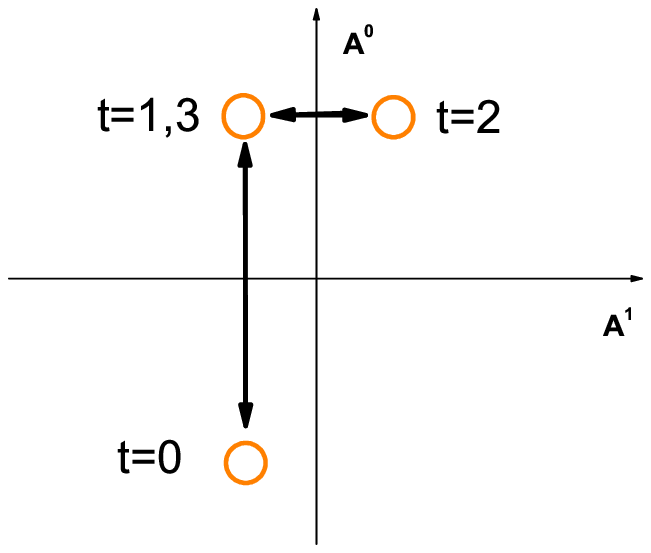,
width=0.53\linewidth}}} \vspace{-0.5cm}
\caption{\label{layoutanalysis}Attractor dynamics when (a)
$|\Delta\textit{A}^{1}|$ is larger and (b) $|\Delta\textit{A}^{0}|$
is larger.}
\end{figure}

Now suppose $\mu=0$ at $t=0$. Following the dynamics sketched in
Fig.~\ref{layoutanalysis}, we can calculate $k_{\mu}$ and
$k_{\overline{\mu}}$ as follows:
\begin{eqnarray}
&&k_{0}(1)=k_{0}(0)-\sqrt{N}A^{0}(0),\hspace{0.3cm}k_{1}(1)=k_{1}(0)\nonumber\\
&&k_{0}(2)=k_{0}(1),\hspace{0.3cm}k_{1}(2)=k_{1}(1)-\sqrt{N}A^{1}(1)\nonumber\\
&&k_{0}(3)=k_{0}(2),\hspace{0.3cm}k_{1}(3)=k_{1}(2)-\sqrt{N}A^{1}(2)\nonumber\\
&&k_{0}(0)=k_{0}(3)-\sqrt{N}A^{0}(3),\hspace{0.3cm}k_{1}(0)=k_{1}(3)\label{m}
\end{eqnarray}



Eqs.~(\ref{deltaA13}-\ref{m}) are the equations of motion in the phase space. When
$A^{1},A^{0}\sim\textit{N}^{-1/2}$ and $k_{1},k_{0}\sim\textit{N}^{0}$, we obtain Eq.~(\ref{step}) for linear
payoffs using Taylor expansion. On the other hand, there exist solutions with non-vanishing step sizes along one
of the two directions, and vanishing step sizes along the remaining direction. Suppose
$\Delta\textit{A}^{0}\sim\textit{N}^{-1/2}$ and $k_{0}(t)\approx\textit{k}_{0}=\textrm{constant}$, and
$\Delta\textit{A}^{1}\sim\textit{N}^{0}.$ From $k_{1}(0)=k_{1}(3)$ in Eq.~(\ref{m}), we can get
$\sqrt{N}A^{1}(1)+\sqrt{N}A^{1}(2)=0$. Since $A^{1}(2)=A^{1}(1)+\Delta\textit{A}^{1}(1)$, we have
$$\textit{A}^{1}(1)=-\frac{1}{2}\Delta\textit{A}^{1}(1)$$ and
$$\textit{A}^{1}(2)=\frac{1}{2}\Delta\textit{A}^{1}(1)$$.

Let $k_{1}(0)=k_{1}(1)=k_{1}(3)=\overline{k_{1}}-\sqrt{N}\Delta\textit{A}^{1}(1)/4$ and
$k_{1}(2)=\overline{k_{1}}+\sqrt{N}\Delta\textit{A}^{1}(1)/4$, where $\overline{k_{1}}$ is the average value of
$k_{1}(t)$. By Eq.~(\ref{deltaA13}),
\begin{eqnarray}
&&\langle\Delta\textit{A}^{1}(1)\rangle=\frac{1}{8}\Biggl[\mathrm{erf}\frac{\Delta\textit{A}^{1}(1)+4(\overline{k_{1}}+k_{0})/\sqrt{N}}{\sqrt{8\rho}}\nonumber\\
&&+\mathrm{erf}\frac{\Delta\textit{A}^{1}(1)-4(\overline{k_{1}}+k_{0})/\sqrt{N}}{\sqrt{8\rho}}+2\mathrm{erf}\frac{\Delta\textit{A}^{1}(1)+4\overline{k_{1}}/\sqrt{N}}{\sqrt{8\rho}}\nonumber\\
&&+2\mathrm{erf}\frac{\Delta\textit{A}^{1}(1)-4\overline{k_{1}}/\sqrt{N}}{\sqrt{8\rho}}+\mathrm{erf}\frac{\Delta\textit{A}^{1}(1)+4(\overline{k_{1}}-k_{0})/\sqrt{N}}{\sqrt{8\rho}}\nonumber\\
&&+\mathrm{erf}\frac{\Delta\textit{A}^{1}(1)-4(\overline{k_{1}}-k_{0})/\sqrt{N}}{\sqrt{8\rho}}\Biggr].
\end{eqnarray}

In general, the dynamics converges with $k_{1},k_{0}\sim\textit{N}^{0}$, yielding the self-consistent equation
for $\Delta\textit{A}^{1}$ in Eq.~(\ref{32}) (we have made the dependence on $t=1$ implicit).

To find the variance, we use Eq.~(\ref{sigma2}). Knowing
$\langle\textit{A}^{\mu^{*}(t)}\rangle=[A^{0}(0)+A^{1}(1)+A^{1}(2)+A^{0}(3)]/4=0,$ the variance is calculated
as:
$$\frac{\sigma^{2}}{N}=\frac{N}{16}[(-\frac{1}{2}\Delta\textit{A}^{1}(1))^{2}+(\frac{1}{2}\Delta\textit{A}^{1}(1))^{2}],$$
resulting in the expression for $\sigma^{2}/N$ in Eq.~(\ref{32}).

\section{The stability of the step size in the secondary direction}

We consider the stability of the step size in the secondary direction. Following the notation in Appendix A, we
assume $(\Delta\mathrm{A}^{1}(1)\sim\textit{N}^{0})$ and $\mu=0$ at $t=0$, and the second direction is $A^{0}$.

Suppose there is a small perturbation $\delta\mathrm{A}^{0}(0)$ at
$t=0$, then
$$k_{0}(1)=k_{0}(0)-\sqrt{N}\delta\textit{A}^{0}(0)=-\sqrt{N}\delta\textit{A}^{0}(0).$$
$$k_{1}(1)=k_{1}(0)=\overline{k_{1}}-\frac{\sqrt{N}}{4}\Delta\textit{A}^{1}(1)\approx-\frac{\sqrt{N}}{4}\Delta\textit{A}^{1}(1).$$

We obtain
\begin{eqnarray}
&&\Delta\textit{A}^{0}(0)=\frac{1}{8}[-\mathrm{erf}\frac{2\delta\textit{A}^{0}(0)+\frac{1}{2}\Delta\textit{A}^{1}(1)}{\sqrt{2\rho}}\nonumber\\
&&-2\mathrm{erf}\frac{2\delta\textit{A}^{0}(0)}{\sqrt{2\rho}}-\mathrm{erf}\frac{2\delta\textit{A}^{0}(0)-\frac{1}{2}\Delta\textit{A}^{1}(1)}{\sqrt{2\rho}}].
\end{eqnarray}

Using Taylor expansion to the first order,
\begin{eqnarray}
&&\Delta\textit{A}^{0}(0)=\nonumber\\
&&-\frac{1}{\sqrt{2\pi\rho}}(1+e^{-\frac{(\Delta\textit{A}^{1}(1))^{2}}{8\rho}})\delta\textit{A}^{0}(0)\equiv\delta\textit{A}^{0}(1).\nonumber\\
&&\Delta\textit{A}^{1}(0)=0.
\end{eqnarray}

Similarly, for $t=1$,
\begin{eqnarray}
&&k_{0}(2)=-\sqrt{N}\delta\textit{A}^{0}(0), \quad\textit{k}_{1}(2)=\frac{\sqrt{N}}{4}\Delta\textit{A}^{1}(1)\nonumber\\
&&\Delta\textit{A}^{1}(1)=\mathrm{erf}(\frac{\Delta\textit{A}^{1}(1)}{\sqrt{8\rho}}),\quad\Delta\textit{A}^{0}(1)=0.
\end{eqnarray}
for $t=2$,
\begin{eqnarray}
&&k_{0}(3)=-\sqrt{N}\delta\textit{A}^{0}(0), \quad\textit{k}_{1}(3)=-\frac{\sqrt{N}}{4}\Delta\textit{A}^{1}(1)\nonumber\\
&&\Delta\textit{A}^{1}(2)=-\mathrm{erf}(\frac{\Delta\textit{A}^{1}(1)}{\sqrt{8\rho}}),\quad\Delta\textit{A}^{0}(2)=0.
\end{eqnarray}
for $t=3$,
\begin{eqnarray}
&&k_{0}(4)=-2\sqrt{N}\delta\textit{A}^{0}(0)-\sqrt{N}\delta\textit{A}^{0}(1)\nonumber\\
&&k_{1}(4)=-\frac{\sqrt{N}}{4}\Delta\textit{A}^{1}(1)\nonumber\\
&&\Delta\textit{A}^{0}(3)=-\frac{1}{\sqrt{2\pi\rho}}(1+e^{-\frac{(\Delta\textit{A}^{1}(1))^{2}}{8\rho}})(\delta\textit{A}^{0}(0)+\delta\textit{A}^{0}(1))\nonumber\\
&&\Delta\textit{A}^{1}(3)=0.
\end{eqnarray}
$\delta\textit{A}^{0}(4)$ is the accumulated perturbation from the previous time steps. That is,
\begin{eqnarray}
&&\Delta\textit{A}^{0}(4)=\delta\textit{A}^{0}(0)+\delta\textit{A}^{0}(1)\nonumber\\
&&-\frac{1}{\sqrt{2\pi\rho}}(1+e^{-\frac{(\Delta\textit{A}^{1}(1))^{2}}{8\rho}})(\delta\textit{A}^{0}(0)+\delta\textit{A}^{0}(1)).
\end{eqnarray}
leading to Eq.~(\ref{stability}).

\section{The equations of motion in MG with bimodal preference distribution}
For bimodal distribution of preferences of strategies, the calculation is similar to that of Appendix A, except
changing $P_{\omega_{ia}}$ from $e^{-\omega^{2}/2R}/\sqrt{2\pi\textit{R}}$ to
$\delta(\omega_{ia}-\sqrt{\rho\textit{N}})/2+\delta(\omega_{ia}+\sqrt{\rho\textit{N}})/2$.

For $m=1$, the theoretical result is as follows: Denoting the step functions by
$\Theta_{1}=\Theta(x+\sqrt{\rho\textit{N}})$,$\quad$$\Theta_{2}=\Theta(x-\sqrt{\rho\textit{N}})$, we have
\begin{eqnarray}
&&A_{\mu}=\Theta_{1}(2k_{\mu}(t)+2k_{\overline{\mu}}(t))+\Theta_{2}(2k_{\mu}(t)+2k_{\overline{\mu}}(t))\nonumber\\
&&-\Theta_{1}(-(2k_{\mu}(t)+2k_{\overline{\mu}}(t)))-\Theta_{2}(-(2k_{\mu}(t)+2k_{\overline{\mu}}(t))),\nonumber\\
\nonumber\\
&&B_{\mu}=\Theta_{1}(2k_{\mu}(t))+\Theta_{2}(2k_{\mu}(t))\nonumber\\
&&-\Theta_{1}(-2k_{\mu}(t))-\Theta_{2}(-2k_{\mu}(t)),\nonumber\\
\nonumber\\
&&C_{\mu}=\Theta_{1}(2k_{\mu}(t)-2k_{\overline{\mu}}(t))+\Theta_{2}(2k_{\mu}(t)-2k_{\overline{\mu}}(t))\nonumber\\
&&-\Theta_{1}(-(2k_{\mu}(t)-2k_{\overline{\mu}}(t)))-\Theta_{2}(-(2k_{\mu}(t)-2k_{\overline{\mu}}(t)))\nonumber\\
\end{eqnarray}

For online update
\begin{eqnarray}
&&k_{\mu}(t+1)=k_{\mu}(t)-\sqrt{N}A^{\mu}(t),\nonumber\\
&&k_{\overline{\mu}}(t+1)=k_{\overline{\mu}}(t),\nonumber\\
\end{eqnarray}

then
\begin{eqnarray}
&&\Delta\textit{A}^{\mu}(t)=\frac{1}{16}[A_{\mu}(t+1)+2B_{\mu}(t+1)+C_{\mu}(t+1)\nonumber\\
&&-(A_{\mu}(t)+2B_{\mu}(t)+C_{\mu}(t))].
\end{eqnarray}

\begin{eqnarray}
&&\Delta\textit{A}^{\overline{\mu}}(t)=\frac{1}{16}[A_{\mu}(t+1)-C_{\mu}(t+1)\nonumber\\
&&-(A_{\mu}(t)-C_{\mu}(t))].
\end{eqnarray}

Likewise, for batch update, with the notation $\overline{\mu}=1-\mu$,

\begin{eqnarray}
&&A_{\mu}=\Theta_{1}(2k_{\mu}(t)+2k_{\overline{\mu}}(t))+\Theta_{2}(2k_{\mu}(t)+2k_{\overline{\mu}}(t))\nonumber\\
&&-\Theta_{1}(-(2k_{\mu}(t)+2k_{\overline{\mu}}(t)))-\Theta_{2}(-(2k_{\mu}(t)+2k_{\overline{\mu}}(t))),\nonumber\\
\nonumber\\
&&B_{\mu}=\Theta_{1}(2k_{\mu}(t))+\Theta_{2}(2k_{\mu}(t))\nonumber\\
&&-\Theta_{1}(-2k_{\mu}(t))-\Theta_{2}(-2k_{\mu}(t)),\nonumber\\
\nonumber\\
&&C_{\mu}=\Theta_{1}(2k_{\mu}(t)-2k_{\overline{\mu}}(t))+\Theta_{2}(2k_{\mu}(t)-2k_{\overline{\mu}}(t))\nonumber\\
&&-\Theta_{1}(-(2k_{\mu}(t)-2k_{\overline{\mu}}(t)))-\Theta_{2}(-(2k_{\mu}(t)-2k_{\overline{\mu}}(t)))\nonumber\\
\end{eqnarray}

Since
\begin{eqnarray}
k_{\mu}(t+1)=k_{\mu}(t)-\sqrt{N}A^{\mu}(t),\nonumber\\
\end{eqnarray}

we obtain
\begin{eqnarray}
&&\Delta\textit{A}^{\mu}(t)=\frac{1}{16}[A_{\mu}(t+1)+2B_{\mu}(t+1)+C_{\mu}(t+1)\nonumber\\
&&-(A_{\mu}(t)+2B_{\mu}(t)+C_{\mu}(t))].
\end{eqnarray}

\section*{ACKNOWLEDGEMENTS}
We thank S. W. Lim, and C. H. Yeung for discussions. This work is supported by the Research Grant Council of
Hong Kong (DAG05/06.SC36 and HKUST 603606).


\begin{thebibliography}{0}

\bibitem{Anderson1988}
P. W. Anderson, K. J. Arrow, and D. Pines, {\it The Economy as an
Evolving Complex System} (Addison Wesley, Redwood City, CA, 1988).

\bibitem{Challet1997}
D. Challet, M. Marsili, and Y. C. Zhang, Physica A, {\bf 246}, 407
(1997).

\bibitem{Wei1995}
G. Wei$\beta$ and S. Sen, {\it Adaption and learning in multi-agent
systems, Lecture Notes in Computer Science} {\bf 246} (Springer,
Berlin, 1995).

\bibitem{Schweitzer2002}
F. Schweitzer (ed.), {\it Modeling Complexity in Economic and Social
Systems} (World Scientific, Singapore, 2002).

\bibitem{Challet2000}
D. Challet, M. Marsili, and R. Zecchina, Phys. Rev. Lett. {\bf 84}, 1824 (2000).

\bibitem{Marsili2000}
M. Marsili, D. Challet, and R. Zecchina, Physica A {\bf 280}, 522
(2000).

\bibitem{Heimel2001}
J. A. F. Heimel and A. C. C. Coolen, Phys. Rev. E {\bf 63}, 056121
(2001).

\bibitem{Coolen2001}
A. C. C. Coolen and J. A. F. Heimel, and D. Sherrington, Phys. Rev. E {\bf 65}, 016126 (2001).

\bibitem{Coolen2005}
A. C. C. Coolen, \textit{The Mathematical Theory of Minority Games}
(Oxford University Press, Oxford, UK, 2005).

\bibitem{Marsili2001}
M. Marsili, Physica A {\bf 299}, 93 (2001).

\bibitem{Andersen2003}
J. V. Andersen and D. Sornette, Eur. Phys. J. B {\bf 31}, 141 (2003).

\bibitem{Giardina2003}
I. Giardina and J.-P. Bouchaud, Eur. Phys. J. B {\bf 31}, 421 (2003).

\bibitem{de Martino2004}
A. de Martino, I. Giardina, M. Marsili, and A. Tedeschi,
Phys. Rev. E {\bf 70}, 025104(R) (2004).

\bibitem{Tedeschi2005}
A. Tedeschi, A. De Martino, and I. Giardina, Physica A 358, 529 (2005).

\bibitem{Yeung2007}
C. H. Yeung, K. Y. M. Wong, and Y.-C. Zhang,
arXiv:0708.0209 (2007).

\bibitem{Savit1999}
R. Savit, R. Manuca, and R. Riolo, Phys. Rev. Lett. {\bf 82}, 2203 (1999).

\bibitem{Manuca2000}
R. Manuca, Y. Li, R. Riolo, and R. Savit, Physica A {\bf 282}, 559 (2000).

\bibitem{Li2000}
Y. Li, A. VanDeemen, and R. Savit, Physica A {\bf 284}, 461 (2000).

\bibitem{Lee2003}
K. Lee, P. M. Hui, and N. F. Johnson, Physica A {\bf 321}, 309 (2003).

\bibitem{Wong2004}
K. Y. M. Wong, S. W. Lim, and Z. Gao, Phys. Rev. E {\bf 70}, 025103(R) (2004).

\bibitem{Wong2005}
K. Y. M. Wong, S. W. Lim, and Z. Gao, Phys. Rev. E {\bf 71}, 066103 (2005).

\bibitem{Sherrington2002}
D. Sherrington, E. Moro, and J. P. Garrahan, physica A 311, 527 (2002).

\bibitem{Moro2004}
E. Moro, Advances in Condensed Matter and Statistical Physics, E Korutcheva and R. Cuerno(eds.), (Nova, New
York), 263 (2004).

\bibitem{Garranhan2000}
J. P. Garrahan, E. Moro, and D. Sherrington, Phys. Rev. E 62, R9 (2000).

\bibitem{Ting2004}
Y. S. Ting, Master of Philosophy Thesis, Hong Kong University of
Science and Technology(2004).



\end{thebibliography}
\end{document}